\documentclass[11pt,a4paper]{article}
\usepackage{jheppub}
\usepackage{amsmath,amsthm,mathabx,amssymb}
\usepackage[dvipsnames,svgnames,x11names]{xcolor}

\usepackage{tikz-cd}
\usetikzlibrary{arrows.meta, automata,  
                positioning,
                quotes}
\usetikzlibrary{decorations.markings}
    \tikzset{middlearrow/.style={
        decoration={markings,
            mark= at position 0.75 with {\arrow{#1}} ,
        },
        postaction={decorate}
    }
}

\usepackage[mathscr]{eucal}
\usepackage[final]{showkeys}
\usepackage{esint}
\usepackage{braket}
\usepackage{bbold}
\usepackage{hyperref}
\hypersetup{
    colorlinks=true,                 % false: boxed links; true: colored links
    linkcolor=black!85,                 % color of internal links (change box color with linkbordercolor)
    linktoc=all,                     % link over complete toc entries
    citecolor=SeaGreen,                 % color of links to bibliography
    filecolor=pink,                  % color of file links
    urlcolor=green!35!black!100                 % color of external links
}

\newcommand{\p}{\partial}
\newcommand{\cG}{\mathcal{G}}

\newcommand{\tp}{\mathtt{p}}
\newcommand{\Tr}{\mathrm{Tr}}

\newcommand{\M}{\mathbb{M}}
\newcommand{\PP}{\mathbb{P}}
\newcommand{\C}{\mathbb{C}}
\newcommand{\onebb}{\mathbb{1}}
\newcommand{\nn}{\nonumber}

\newcommand{\sa}{\mathsf{a}}
\newcommand{\cP}{\mathcal{P}}
\newcommand{\cJ}{\mathcal{J}}
\newcommand{\cC}{\mathcal{C}}
\newcommand{\cD}{\mathcal{D}}
\newcommand{\cK}{\mathcal{K}}
\newcommand{\cR}{\mathcal{R}}
\newcommand{\cA}{\mathcal{A}}
\newcommand{\cQ}{\mathcal{Q}}
\newcommand{\cH}{\mathcal{H}}
\newcommand{\cL}{\mathcal{L}}
\newcommand{\cV}{\mathcal{V}}

\newcommand{\sK}{\mathsf{K}}
\newcommand{\sD}{\mathsf{D}}

\newcommand{\tH}{\mathtt{H}}
\newcommand{\ib}{\mathtt{I}}

\newcommand{\tE}{\mathtt{E}}

\newcommand{\Sch}{\mathtt{Sch}}
\newcommand{\tz}{\mathtt{z}}

\newcommand{\sign}{\mathrm{sign}}

\theoremstyle{definition}

\newtheorem*{lemma*}{Lemma}
\newtheorem*{theorem*}{Theorem}
\newtheorem*{lieb*}{Lieb's Concavity Theorem}
\newtheorem*{prop*}{Proposition}

\theoremstyle{remark}

\newcommand{\mso}{\mathfrak{so}}

\title{\center  Three-dimensional non-relativistic chiral massive higher-spin gravity}
\author{Arpita Mitra$^{a}$,}
\author{Debangshu Mukherjee$^{b}$}
\author{and Tung Tran$^{c}$}
\affiliation{$^{a}$Centre for Theoretical Physics and Natural Philosophy,\\
Nakhonsawan Studiorum for Advanced Studies, Mahidol University, Thailand\\ }
\affiliation{$^{b}$Department of Physics, Pohang University of Science \& Technology (POSTECH), \\
77 Cheongam-ro, Nam-gu, Pohang-si, Gyeongsangbuk-do, 37673, Republic of Korea.\\ }
\affiliation{$^{c}$Department of Physics and Research Institute of Basic
  Science, \\ Kyung Hee University, Seoul 02447, Korea.\\ }

\emailAdd{arpita.mit@mahidol.ac.th}
\emailAdd{debangshum@postech.ac.kr}
\emailAdd{tung.tran@khu.ac.kr}

\abstract{We obtain a non-relativistic chiral massive higher-spin gravity in a deformed $AdS_3$ spacetime by applying a Lifshitz deformation and subsequent null reduction to chiral massless higher-spin gravity in 
$AdS_4$. Intriguingly, the vertices of this non-relativistic theory are less constrained than the ones in the original $4d$ chiral massless theory since we do not have enough dynamical generators to fix the couplings uniquely. Anticipating higher-spin interactions should be suppressed, we propose a simple approximate mass-spin relation which interpolates between the relativistic and non-relativistic regimes. With the proposed mass-spin relation, we observe that that higher-spin interactions indeed become suppressed at large spins, consistent with low-energy physics. We conjecture that the holographic dual of the non-relativistic chiral massive higher-spin gravity proposed in this work is a $2d$ non-relativistic Landau-Ginzburg theory in the light-cone gauge. This non-relativistic theory is expected to describe a two-fluid system with a $\lambda$-point constrained in one spatial dimension. }
\begin{document}
\maketitle
%\tableofcontents
% \newpage

%%%%%%%%%%%%%%%%%%%%%%%%%%%%%%%%%%%%%%%%%%%%%%%%%%%%%%%%%%%%%%%%%%%%%%%%%%%%%%%%%%%%%
%%%%%%%%%%%%%%%%%%%%%%%%%%%%%%%%%%%%%%%%%%%%%%%%%%%%%%%%%%%%%%%%%%%%%%%%%%%%%%%%%%%%%

%%%%%%%%%%%%%%%%%%%%%%%%%%%%%%%%%%%%%%%%%%%%%%%%%%%%%%%%%%%%%

%%%%%%%%%%%%%%%%%%%%%%%%%%%%%%%%%%%%%%%%%%%%%%%%%%%%%%%%%%%%%
\section{Introduction}
While studying relativistic theories is essential for probing ultraviolet (UV) physics, it can be equally valuable to find suitable non-relativistic (NR) regimes of the same theories, where the required energy to test computable observables is more accessible. Note that this inevitably contracts the relativistic symmetric group, be it conformal or Poincar\'e, to a smaller subgroup as Lorentz boosts descend to Galilean boosts, leading to different scalings of space and time \cite{Jackiw:1972cb,Hagen:1972pd,Niederer:1972zz} (see also \cite{Duval:2024eod} for a short account on this line of work). These scenarios are known to appear in strongly correlated systems, or condensed matter near criticality, see e.g. \cite{Henkel:1993sg,Son:2005rv,Hoyos:2011ez,Geracie:2015drf} and references therein. (See also \cite{Nastase:2017cxp}.) It is also well-known that obtaining consistent non-relativistic theories from interacting relativistic ones, especially when gravity is involved, is a challenging task, see e.g. \cite{Son:2013rqa,Jensen:2014aia,Geracie:2015dea,Geracie:2015xfa}. Aligning with this line of work is the construction of covariant NR field theories on curved backgrounds by gauging the global Galilean symmetry \cite{Hartong:2014pma,Banerjee:2014nja, Banerjee:2014pya}, and anisotropic scaling symmetry on the flat spacetime \cite{Mitra:2015twa}.

Two of the most common routes in taking the NR limit of relativistic theories are the following\footnote{See e.g. \cite{Banerjee:2018pvs} for a review}:
\begin{itemize}
    \item[1-] The In\"on\"u-Wigner contraction. In this approach, the speed of light $c$ is sent to $\infty$, resulting in some simplification of the relativistic algebra, see e.g. \cite{Hartong:2022lsy} and references therein for a review. As the number of generators remain unchanged in this approach, it permits a natural embedding of generators of the contracted algebra back to the original relativistic algebra, see e.g. \cite{Campoleoni:2021blr}. Note, however, that when taking the Galilean limit of a massless relativistic theory, the notion of mass -- which is crucial and must present in NR systems -- is sometimes not well-defined, see e.g. \cite{Coraddu:2009sb,Jafari:2024bph}.
    
    \item[2-] The second route for taking the NR limit of a relativistic theory is to consider Lifshitz anisotropic scaling, characterized by a dynamical exponent $\tz$, which drives the original relativistic system toward a Schr\"odinger-like regime \cite{Jackiw:1972cb,Hagen:1972pd,Niederer:1972zz}. %In this approach, time $t$ and space $x$ scale differently under Lifshitz scaling, i.e. $t\mapsto \lambda^{\tz}t$ and $x\rightarrow \lambda\,x$, which leads to the notion of absolute time. 
    In this approach, one usually starts with a relativistic theory in $d$ dimensions formulated in the light-cone gauge. Then, upon introducing Lifshitz scalings (which break Lorentz invariance), and performing a compactification along one of the light-cone directions, one can obtain a $(d-1)$-dimensional NR massive theory \cite{Son:2008ye}. (This procedure may be viewed as a null reduction \cite{duval1985bargmann}.) In this case, the masses of the resulting NR fields can be viewed as discrete momentum associated with the contracted light-cone direction. Notably, in this approach, a portion of interacting vertices can remain stable under Lifshitz scaling and dimensional compactification, making it a compelling approach for obtaining consistent non-relativistic systems from relativistic ones.
\end{itemize}
With what presented above, we aim to study the NR limit, via Lifshitz anisotropic scalings, of a specific relativistic model, which is believed to be UV finite (due to an underlying infinite-dimensional symmetry that tends to suppress any UV divergences). This model is chiral higher-spin gravity (HSGRA) \cite{Ponomarev:2016lrm}. It was first discovered in the light-cone gauge based on the works \cite{Bengtsson:1983pd,Metsaev:1991mt,Metsaev:1991nb,Metsaev:2018xip}, and was subsequently covariantized in e.g. \cite{Sharapov:2022faa,Sharapov:2022awp} in terms of equations of motion. (See also \cite{Ponomarev:2017nrr,Monteiro:2022xwq} for various subsectors of chiral HSGRA in spacetime, and \cite{Krasnov:2021nsq,Herfray:2022prf,Adamo:2022lah,Neiman:2024vit,Tran:2025uad} for their twistor duals.)\footnote{See further \cite{Ivanovskiy:2023aay,Ivanovskiy:2025ial,Serrani:2025owx,Serrani:2025oaw} for recent development related to chiral HSGRA in the light-cone gauge. See also \cite{Metsaev:2022yvb,Metsaev:2022ndg,Ponomarev:2022vjb}.}

Several checks performed at one loop in spacetime \cite{Skvortsov:2018jea,Skvortsov:2020wtf,Skvortsov:2020gpn} and at two loops using twistor technique in \cite{Tran:2025xbt} suggest that chiral HSGRA is likely UV finite to all orders in perturbation theory. Intriguingly, the holographic dual of chiral HSGRA in $AdS_4$ is anticipated to be the chiral sector of the well-known $3d$ Chern-Simons matter theories at large $N$, see \cite{Jain:2024bza,Aharony:2024nqs}. Due to the strong integrability of both the gravitational and conformal field theories, the duality is believed to be exact to all orders in perturbation theory. It is then natural to ask whether the above exact duality between the pair of theories mentioned above can be continuously deformed into a lower-dimensional NR massive dual pair.

In this work, we begin our exploration of the above proposal from the gravity side, i.e. with chiral HSGRA in $AdS_4$. In particular, we will introduce a Lifshitz anisotropy scaling to drift the $4d$ massless theory to a $3d$ NR massive one. Due to the fact that Schr\"odinger algebra is easily realized in light-cone coordinates, it is rather suggestive to start with a light-cone description of chiral HSGRA in $AdS_4$. Fortunately, most of the work has already been done in \cite{Metsaev:2018xip,Skvortsov:2018uru} up to the cubic order. (See also \cite{Lang:2025rxt} for another light-cone construction for a higher-spin extension of self-dual Yang-Mills and self-dual gravity in $dS_4$.) As a result, our task is relatively simple -- we only need to analyze this gravitational theory in the NR limit. 

Our work is organized as follows. Section \ref{sec:2} provides some relevant background on Lifshitz anisotropic scaling, Schr\"odinger algebra, as well as a crash course on light-front formalism for massless fields in $AdS_4$. Section \ref{sec:3} studies a specific Schr\"odinger geometry derived from exact $AdS_4$ in the light-cone gauge, which is twist-free and torsional. These materials compose the first part of the paper.

In the second part, we shall explore some simple non-relativistic dynamics after Lifshitz scalings being introduced to break Lorentz invariance in Section \ref{sec:NR-LC}. Upon doing a dimensional compactification along one of the directions of the light cone to obtain a NR chiral massive HSGRA in $3d$ in Section \ref{sec:theory}. Some simple correlation functions, and the effect of higher-spin interaction suppression are also studied in the same section. Collecting the insights we gain from this work, we propose a conjecture about the duality between the obtained NR $3d$ chiral massive HSGRA with a certain NR chiral massive critical weakly coupled scalar theory in $(1+1)$ dimensions in Section \ref{sec:holo}. We expect that if our conjecture is true (up to certain degree), it may be relevant in the context of two-fluid system constrained in one spatial dimension. We conclude with some remarks in Section \ref{sec:diss}. There are also two appendices which the reader can refer to in due course.

%\paragraph{Note on changes.} Please indicate the big changes in shuffling material here. 
%\begin{enumerate}
%    \item Originally Sec 4.2. had some details of EOM of Schr\"{o}dinger and its solutions. Now, it just talks about the bulk-to-boundary propagator. All details of EOM, explicit computation of the propagator and 2-point function goes in Appendix \ref{app:2pt}.
%\end{enumerate}
%%%%%%%%%%%%%%%%%%%%%%%%%%%%%%%%%%%%%%%%%%%%%%%%%%%%%%%%%%%%%
\section{Review}\label{sec:2}
This section reviews some basic ingredients to study a Lifshitz deformation of chiral higher-spin gravity in $AdS_4$. Our convention aligns with those in \cite{Metsaev:1999ui}.

%%%%%%%%%%%%%%%%%%%%%%%%%%%%%%%%%%%%%%%%%%%%%%%%%
\subsection{Lifshitz anisotropic scaling and Schr\"odinger algebra}\label{sec:Lif-Sch-alg}
%We begin with a brief review of Schr\"odinger algebras which includes Lifshitz anisotropic scalings. 

It is well-known that the conformal algebra of a $(d+2)$-dimensional flat spacetime $\M_{d+2}$ is $\mso(d+2,2)$, which happens to coincide with the isometry algebra of $AdS_{d+3}$. Within this Lie algebra, one can look for a subalgebra known as the Schr\"odinger algebra $\Sch_{\tz}(d)$ parametrized by a Lifshitz dynamical exponent $\tz$ and the number of spatial dimensions $d$. This is the algebra that is relevant for studying various condensed matter systems, cf. \cite{Henkel:1993sg,Son:2005rv,Hoyos:2011ez,Geracie:2015drf}. 

Let $(\cP^A,\cJ^{AB},\cD,\cK^A)$ be the generators corresponding to translations, Lorentz rotations, dilatation and special conformal boosts of the conformal algebra $\mso(d+2,2)$ where $A=0,1,\ldots,d+1$. Then, the generators 
\begin{align}
    (P^+,P^-,P^a,J^{ab},G^a,D,K)\,,\qquad a=1,\ldots,d\,,
\end{align}
of the Schr\"odinger algebra $\Sch_{\tz}(d)$ can be obtained from $(\cP^A,\cJ^{AB},\cD,\cK^A)$ as (see e.g. \cite{Bergshoeff:2014uea})
\begin{subequations}\label{eq:Bergshoeff-convention}
    \begin{align}
        P^+&=\frac{1}{\sqrt{2}}\big(\cP^{d+1}+\cP^0\big)\,,\qquad &P^-&=\frac{1}{\sqrt{2}}\big(\cP^{d+1}-\cP^0\big)\,,\label{eq:H-N}\\
        D&=(\tz-1)\cJ^{d+1\,0}+\cD\,,\qquad &K&=\frac{1}{\sqrt{2}}(\cK^{d+1}+\cK^0)\equiv \cK^+\,,\\
        G^a&=\frac{1}{\sqrt{2}}\big(\cJ^{d+1\,a}+\cJ^{0\,a}\big)\equiv \cJ^{+a}\,.
    \end{align}
\end{subequations}
Thus, $\Sch_{\tz}(d)$ is indeed a subalgebra of $\mso(d+2,2)$. 

In the literature, see e.g. \cite{Son:2008ye}, $P^+$ is usually referred to as the number or mass generator. The generators $G^a$ are known as the Galilean boosts. They are the descendants of Lorentz boost generators $\cJ^{+a}$ in the light-cone gauge where $x^{\pm}=\frac{1}{\sqrt{2}}(x^{d+1}\pm x^0)$. Note that $P^-$ serves as the Hamiltonian, generating the evolution of the system where $\Sch_{\tz}(d)$ is the underlying symmetry.

What differs $\Sch_{\tz}(d)$ with other In\"on\"u-Wigner subalgebras of the conformal algebra is the dilatation generator $D$. This generator is special since it is defined with a Lifshitz twist, telling the scaling dimensions of other generators in the Schr\"odinger regime. 

Note that when $\tz \neq 2$, the Schr\"odinger algebra 
does not contain any of the special conformal boost generators $\cK^A$ of $\mso(d+2,2)$ among its generators. On the other hand, when the Lifshitz scaling $\tz=2$, the $\Sch_{\tz=2}(d)$ algebra acquires one special conformal boost generator, which is $\cK^+\equiv K$. In this case, $\{P^-,D,K\}$ form an $\mathfrak{sl}(2,\mathbb{R})$ subalgebra where $[D,P^-]=-P^-$, $[D,K]=2 K$, and $[P^-,K]=D$. For more details, see \cite{Son:2008ye}. %\eqref{eq:commutation-relations-Schrodinger}. 

In this work, the Schr\"odinger algebra of interest will be $\Sch_{\tz=2}(1)\subset \mso(3,2)$. This algebra has six generators, which are $(P^+,P^-,P^1,G^1,D,K)$ -- i.e. no spatial rotations. Since $\Sch_{\tz}(d)$ is conveniently realized in light-cone coordinates, field theory representations built upon $\Sch_{\tz}(d)$, when available, are most naturally formulated in the light-cone gauge. It must be noted that, following the terminology in \cite{Taylor:2015glc}, we shall refer to $\Sch_{\tz}(1)$ as Schr\"{o}dinger algebra for arbitrary $\tz$, even though, in other literature, this terminology is typically reserved only for $\tz=2$. Note that most of our subsequent analysis will, in fact, focus on the case $\tz=2$.

%%%%%%%%%%%%%%%%%%%%%%%%%%%%%%%%%%%%%%%%%%%%%%%%%
\subsection{Light-front formalism for massless fields in \texorpdfstring{$AdS_4$}{AdS}}\label{sec:light-front}

The line element of exact $AdS_4$ spacetime has a particular nice representation in the Poincar\'e patch where
\begin{align}\label{eq:exact-AdS}
    ds^2=\frac{1}{r^2}\Big(-dx_0^2+dx_1^2+dx_2^2+dr^2\Big)=\frac{1}{r^2}\Big(2dx^+dx^-+2dx_1^2+dr^2\Big)\,.
\end{align}
From \eqref{eq:exact-AdS}, one can find two groups of isometry generators: (i) kinematical generators, (ii) dynamical generators, cf. \cite{Metsaev:1999ui,Metsaev:2008fs}, as
\begin{subequations}
    \begin{align}
        \mathrm{\underline{kinematical  \ generators}}\quad &:\qquad P^+,P^1,J^{+1},J^{+-},D,K^+,K^1 
        \,,\\
        \mathrm{\underline{dynamical\  generators}}\quad &:\qquad P^-,J^{-1},K^-
        \,.
    \end{align}
\end{subequations}
On the one hand, the kinematic generators, which do not receive non-linear corrections on the hypersurface $x^+=0$ -- cf. \cite{Metsaev:1999ui} -- play the role of ``constrainers'' that shape the general expressions of the dynamical generators $(P^-,J^{-1},K^-)$. The dynamical generators, on the other hand, are crucial in fixing couplings, thus defining a theory in the light-cone gauge. 

Note that when we specify $x^+$ as the time direction, the number of dynamical generators turn out to be minimal. As a result, the process of studying deformation of the algebra can be slightly simpler (in good cases) compared to covariant approaches. This is one of the advantages of the light-front formalism. (See further general discussion in e.g. \cite{Metsaev:2005ar} and \cite{Ponomarev:2022vjb}.) Let us elaborate this idea in more detail. 

Denoting $\varphi_{\pm s}(x^+,x^-,x^1,r)$ as massless spinning fields in position space, with helicity $\pm s$. These fields can be related to the ones in momentum space via Fourier transform as
\begin{align}\label{eq:fieldFourier}
    \varphi_{\pm s}(x^+,x^-,x^1,r)=\int \frac{d \rho \,d\beta}{2\pi}e^{i(\beta x^-+\rho x^1)}\varphi_{\pm s}(x^+,\beta,\rho,r)\,,\qquad \rho \equiv p^1\,,\qquad \beta\equiv p^+\,.
\end{align}
Note that $\varphi^{\dagger}_{+s}(x^+,\tp,r)=\varphi_{-s}(x^+,-\tp,r)$, where $\tp=(\beta,\rho)$. (It is well-known that, in $4d$, any massless spinning field can be described by two complex scalar fields $\varphi_{\pm s}$, which are hermitian conjugate of each other.) 

At the level of free fields, the charges $Q$ associated with AdS isometry generators $\cG$, which are quadratic in fields, can be defined as:
\begin{align}
    Q=\int  d^2\tp \,dr\,\beta\,\Tr\Big[(\varphi_h)^{\dagger} \cG(x^+,\tp,\p_{\tp},r,\p_r)\varphi_h\Big]\,.
\end{align}
Note that the kinematic generators, which keeps the $x^+=0$ hypersurface invariant, have the following differential expression, cf. \cite{Metsaev:2018xip}, 
\begin{subequations}\label{eq:kinetic-AdS}
    \begin{align}
        P^+&=\beta\,, &P^1&=\rho, \\  J^{+1}&= {i} x^+ \rho + \partial_{\rho}\beta, &J^{+-}&= { i}x^+P^- + \partial_{\beta} \beta\,,\\
        D &= {i} x^+ P^- -\partial_\beta \beta - \partial_{\rho}\rho +r\partial_r + 1\,,\\ 
        K^+&= \frac{1}{2}(2{i} x^+ \partial_\beta
- \partial_{\rho}^2 +r^2)\beta +  {i} x^+ D\,,\\
K^1 &= \frac{1}{2} (2i x^+ \partial_\beta - \partial_{\rho}^2 +r^2)\rho
- \partial_{\rho} D  + M^{r1} r + {i} M^{1-}x^+\,.
\end{align}
\end{subequations}
On the other hand, one finds that the dynamical generators are of the form
\begin{subequations}\label{eq:dynamic-AdS}
    \begin{align}
  P^- &= - \frac{\rho^2}{2\beta} + \frac{\partial_r^2}{2\beta}\,,\\  J^{-1} &= -\partial_\beta \rho + \partial_{\rho} P^- - M^{r1}\frac{1}{\beta}\partial_r \,,\\ 
  K^- &= \frac{1}{2} (2 i  x^+ \partial_\beta - \partial_{\rho}^2 +r^2)P^-
- \partial_\beta D  + \frac{1}{\beta}(\partial_r\partial_{\rho} -r\rho) M^{r1} -\frac{1}{\beta}M^{r1}M^{r1}\,.
\end{align}
\end{subequations}
In working in four dimensions, we can replace the helicity generator $M^{r1}$ by a label $h$. This label stands for the helicity of the field, which $M^{r1}$ acts on. This stems from the fact that the little group associated with massless fields of the isometry group of $AdS_4$ is $O(2)\simeq U(1)$. The helicity $h$ is then simply a convenient label for the $U(1)$ charges. In terms of oscillators, the helicity operator $M^{r1}$ can be realized as (see \cite{Metsaev:2018xip})
\begin{align}\label{eq:helicity-operator}
    M^{r1}=M^{LR}=(\sa^{\dagger})^R\sa^L-(\sa^{\dagger})^L\sa^R\,,\qquad\text{where}\qquad  [\sa^R,(\sa^{\dagger})^L]=1\,,\quad [\sa^L,(\sa^{\dagger})^R]=1\,.
\end{align}
Here, $\sa^{L,R}$ are annihilation operators that act on the vacuum as $\sa_{L,R}|0\rangle=0$. On the other hand, $(\sa^{\dagger})^{L,R}$ are creation operators that generate other higher-spin states as
\begin{align}\label{eq:ket-HS}
    |\varphi_s(\tp,r,\sa)\rangle=\frac{1}{\sqrt{s!}}\Big((\sa^{\dagger}_L)^s\varphi_s(\tp,r)+(\sa^{\dagger}_R)^s\varphi_{-s}(\tp,r)\Big)|0\rangle\,.
\end{align}
Note that $L$ (left) and $R$ (right) are the directions combined from the usual $(r,1)$ directions. Henceforth, we will not use this representation of $M^{r1}$ but the helicity label $h$ for simplicity. %fields associated with $\sa^{\dagger}_L$ has positive helicity, while fields associated with $\sa^{\dagger}_R$ have negative helicity as can be seen from \eqref{eq:ket-HS}.

A crucial step in the light-front approach is to set $x^+=0$, which prevents the kinematic generators from acquiring non-linear corrections in fields. Consequently, the generators that depend on $x^+$ get simplified to:
\begin{subequations}\label{eq:kinetic-AdS-1}
    \begin{align}
        %P^+&=\beta\,, %&P^1&=\rho,
          J^{+1}&=  \partial_{\rho}\beta, &J^{+-}&=  \partial_{\beta} \beta\,,\label{eq:Jpm}\\
        D &= -\partial_\beta \beta - \partial_{\rho}\rho +r\partial_r + 1\,,\\ 
        K^+&= \frac{1}{2}\beta(
- \partial_{\rho}^2 +r^2) \,,\\
K^1 &= \frac{1}{2} (- \partial_{\rho}^2 +r^2)\rho
- \partial_{\rho} D  + h \,r \,,\\ K^- &= \frac{1}{2} ( - \partial_{\rho}^2 +r^2)P^-
- \partial_\beta D  + \frac{1}{\beta}(\partial_r\partial_{\rho} -r\rho)h -\frac{1}{\beta}h^2\,.
\end{align}
\end{subequations}
Denote $\boldsymbol{\sK}=\{P^+,P^1,J^{+1},J^{+-},D,K^+,K^1\}$ to be the set of kinematic generators above. One finds through various relations
\begin{align}
    [\boldsymbol{\sK},\boldsymbol{\sD}]=\boldsymbol{\sK}\,,\qquad [\boldsymbol{\sK},\boldsymbol{\sD}]=\boldsymbol{\sD}\,,
\end{align}
where $[\,,]$ denotes the Poisson-Dirac commutation relations between fields \cite{Metsaev:2018xip}:
\begin{align}\label{eq:Poisson-Dirac}
    \big[\varphi_h(p,r),\varphi_{h'}(p',r')\big]\Big|_{\text{equal }x^+}=\frac{\delta^2(\tp+\tp')\delta(r-r')}{2\beta}\delta_{h+h',0}\,,
\end{align}
and there will be enough constraints to \emph{partially} fix the generic forms of the dynamical generators 
\begin{align}
    \boldsymbol{\sD}=\sum_{n=2}^{\infty}\sD_n\,,
\end{align}
which can be higher order in fields indicated by the subscript $n$ -- the number of fields that enter the vertices (for instance $D_2=\{P^-,J^{-1},K^-\}$). Then, to fix a theory uniquely, one only needs to solve the following set of equations
\begin{align}\label{eq:main-light-front}
    [\boldsymbol{\sD},\boldsymbol{\sD}]=0\,.
\end{align}
This is the general strategy employed in the light-front approach to construct theories from the bottom up.

%%%%%%%%%%%%%%%%%%%%%%%%%%%%%%%%%%%%%%%%%%%%%%%%%%%%%%%%%%%%%
\section{Twist-free torsional Schr\"odinger geometry}\label{sec:3}
This section studies the twist-free torsional Schr\"odinger geometry, cf. \cite{Bergshoeff:2014uea}, resulting from a Lifshitz deformation (aka. null deformation) of the $AdS_4$ metric \eqref{eq:exact-AdS}. This is a type of deformation, that breaks Lorentz invariance such that time will scale differently from space. This type of anisotropic scaling is often observed in condensed matter system in the vicinity of a fixed point. Our goal is to revisit this symmetry geometrically in a gravity theory.

As discussed in \cite{Balasubramanian:2008dm,Balasubramanian:2010uk, Korovin:2013bua}, one can deform $AdS_4$ by introducing the following Lifshitz anisotropic scalings
\begin{align}\label{eq:M-scale}
    x^+\rightarrow \lambda^{\tz} x^+\,,\qquad  x^-\rightarrow \lambda^{2-\tz} x^-\,,\qquad  x_1 \rightarrow \lambda x_1\,,\qquad r \rightarrow \lambda r \,.
\end{align} 
This leads to the following deformed $AdS_4$ line element \cite{Kachru:2008yh,Balasubramanian:2008dm, Taylor:2015glc} (see also \cite{Korovin:2013bua})
\begin{align}\label{eq:AdS-metric-2}
    ds^2= L^2\Big(-\frac{\upsilon^2(dx^+)^2}{r^{2\tz}}+\frac{1}{r^2}\left(2dx^{+}dx^{-}+dx_1^2 +dr^2\right)\Big)\,,
\end{align}
where $L$ is the anti-de Sitter (AdS) radius. (Note that we put $L$ to keep track of the dimensions of generators. It will be set to 1, when everything is clear as day.) Furthermore, $\upsilon$ in \eqref{eq:AdS-metric-2} is an auxiliary parameter of dimension $[\text{length}]^0$, which should be set to $0$ when $\tz=1$, and $1$ when $\tz\in (1,2]$. Notice that the Lifshitz dynamical exponent is geometrized by the radial coordinate $r$. 

Before proceeding, we shall emphasize that away from $\tz=1$, eq. \eqref{eq:AdS-metric-2} is not diffeomorphic to \eqref{eq:exact-AdS}. This stems from the fact that Lifshitz scaling is a global symmetry, rather than a gauge redundancy. In fact, \eqref{eq:AdS-metric-2} can be viewed as a null deformation of $AdS_4$, where the $(dx^+)^2$ term can be viewed as the backreaction on the geometry when introducing the Lifshitz scaling \eqref{eq:M-scale}, cf. \cite{Guica:2010sw}. It breaks Lorentz symmetry explicitly, but, nonetheless, preserves the Schr\"odinger subalgebra $\Sch_{\tz=2}(1)$. %With this null deformation, time and space become two distinct entities. 
Therefore, \eqref{eq:AdS-metric-2} can help to understand Schr\"odinger algebra geometrically.\footnote{In contrast to the usual Lifshitz symmetry, here $x^-$ also scales, cf. \eqref{eq:M-scale}.}

%%%%%%%%%%%%%%%%%%%%%%%%%%%%%%%%%%%%%%%%%%%%%%%%%%%%%%%%%%%%%%% 
\paragraph{Local frame:} Denote $A,B=+,-,1,r$ as tangent space indices. The scalar product on the tangent space reads
\begin{align}
    \eta_{AB}X^AX^B=2X^+X^-+(X^1)^2+(X^r)^2\,.
\end{align}
Here, we avoid introducing the $\eta_{++}$ component in the tangent space, despite the presence of the component $g_{++}$ in \eqref{eq:AdS-metric-2}, as it may lead to ambiguities when raising and lowering tangent space indices. 

To study a twist-free torsional Schr\"odinger geometry, we need to find a suitable set of local frames. %, which are in general degenerate \cite{Bergshoeff:2014uea}. 
A nice set of local frames that naturally encode Lifshitz scalings \eqref{eq:M-scale} are\footnote{There is also another illuminating choice for the local frame such as
\begin{equation}
    e^+_{\ \ \mu} = \frac{L}{r}\eta^{+}_{\ \ \mu} - \frac{L\,\upsilon^2}{2r^{2\tz-1}}\delta^{+}_{\ \ \mu}\, ,\qquad e^{-}_{\ \ \mu}= \frac{L}{r}\eta^{-}_{\ \ \mu}\, ,\qquad e_{\ \mu}^i = \frac{L}{r}\eta_{\ \mu}^i\,.%\qquad i=1,r\,\ .
\end{equation} 
%In the above, the upper indices denote tangent space indices while the lower ones are spacetime indices. 
%This choice of frame fields is particularly illuminating because of the following reason. Had we been working with exact AdS, the desired vielbiens would have been $e^A_{\mu}=\frac{1}{r}\eta^A_{\mu}$. Due to the deformation, since only 
This choice of local frames can single out the $g_{++}$ component of the deformed spacetime metric \eqref{eq:AdS-metric-2}. Indeed, it is easily seen that only the $e^{+}{}_{\mu}$ component gets deformed, while the remainders describe exact $AdS_4$.
}
\begin{align}\label{eq:vierbein}
    e^+{}_{\mu}=\frac{L}{r^{\tz}}\eta^+{}_{\mu}-\frac{L\,\upsilon^2}{2r^{3\tz-2}}\delta^+{}_{\mu}\,,\qquad e^-{}_{\mu}=\frac{L}{r^{2-\tz}}\eta^-{}_{\mu}\,,\qquad e^i{}_{\mu}=\frac{L}{r}\eta^i{}_{\mu}\,, \qquad i=1,r\,,
\end{align}
where $\mu=+,-,1,r$ denote spacetime indices. In working with the vierbein, our convention is such that the tangent indices will be on the left, while the spacetime indices will be on the right. We raise and lower tangent indices with $\eta_{AB}$, while the same actions for spacetime indices will be done with $g_{\mu\nu}$. Note that even though the metric is not degenerate, there is a slight degeneracy when converting spacetime indices to tangent space indices and vice versa. (This is an unavoidable feature due to the way we define $\eta_{AB}$.) Nonetheless, we find the following set of inverse vierbein
\begin{equation}
    E_-{}^{\mu}=\frac{r^{2-\tz}}{L}\eta_-{}^{\mu}+\frac{\upsilon^2 r^{4-3\tz}}{2L}\delta_-{}^{\mu}\ ,\qquad E_+{}^{\mu}=\frac{r^{\tz}}{L}\eta_+{}^{\mu}\,,\qquad E_i{}^{\mu}=\frac{r}{L}\,\eta_i{}^{\mu}\ ,
\end{equation}
via $ E^{A\mu}e^B{}_{\mu}=\eta^{AB}$. It can be checked that $E_{A}{}^{\nu}e^A{}_{\mu}=\delta^{\nu}{}_{\mu}$\ , along with 
\begin{align}
 E^{A\mu}e^B{}_{\mu}=\eta^{AB}\,,\qquad E^{A\mu}E^{B\nu}\eta_{AB}= g^{\mu\nu}\,.
\end{align}
Notice that the above defining vierbein and their inverses for the geometry \eqref{eq:AdS-metric-2} do not lead to the same Newton-Cartan geometry found in \cite{Bergshoeff:2014uea}, since we have $E^{A\mu}E^{B\nu}\eta_{AB}= g^{\mu\nu}$.
Furthermore, 
%However, we can define the projective vierbein:
%\begin{align}
    %E^{+\mu}=\frac{1}{L}\eta^{+\mu}\,,\qquad 
%    E^{-\mu}=\frac{1}{L}r^{\tz}\delta^{-\mu}\,,\qquad E^{i\mu}=\frac{1}{L}r\,\eta^{i\mu}\,,
%\end{align}
as $g^{i\mu}e^+{}_{\mu}=0$, $e^+{}_{\mu}$ is orthogonal to the spatial direction and defines the so-called \emph{``clock''} form. 
Henceforth, whenever we have to deal with the inverse vierbein of Schr\"odinger geometry explicitly, we shall implicitly assume $\tz=2$. 

%%%%%%%%%%%%%%%%%%%%%%%%%%%%%%%%%%%%%

\paragraph{Gauging Schr\"odinger symmetry:} 

From the line element \eqref{eq:AdS-metric-2}, one can read off the following generators \cite{Metsaev:1999ui} 
in terms of light-cone momenta $P^+$ and transverse momenta $P^1$, 
\begin{subequations}\label{eq:Sch-gen}
    \begin{align}
       \underline{\text{undeformed}}&:   &P^+&=\beta\,,\quad \ \ P^1=\rho\,,\quad \ \ G^1=\p_{\rho}\beta\,,\ \ \ \ \ K^+=-\frac{1}{2}\beta\big(\p_{\rho}^2-r^2\big)\,,\label{eq:undeformed}\\
       \underline{\text{deformed}}&: &\tilde{D}&=-(2-\tz)\p_{\beta}\beta-\p_{\rho}\rho+r\p_r+1\,, \\
       &\qquad &P^- &= - \frac{\rho^2}{2\beta} -\frac{\upsilon^2r^{2-2\tz}}{2%L^2
       }\beta+\frac{\partial_r^2}{2\beta}\,,\label{eq:Hamiltonian}
    \end{align}
\end{subequations}
\normalsize
on the hypersurface $x^+=0$. Note that these are differential operators that \emph{act} on the field representations introduced in \eqref{eq:fieldFourier}. 
Here, we treat $x^+$ as time and $\partial^-$ as time derivatives throughout.

As alluded to above, there will be 6 generators in $\Sch_{\tz}(1)$ at $\tz=2$. Here, we show explicitly how the auxiliary parameter $\upsilon$ enters $P^-$ for dimensional reasons. (It may be set to 1 if there is no ambiguity.) From the scaling \eqref{eq:M-scale}, it is clear that $P$ generators scale as
\begin{align}
    P^-\rightarrow\lambda^{-\tz}P^-\,,\quad P^+\rightarrow \lambda^{\tz-2}P^+,\quad P^1\rightarrow \lambda^{-1}P^1\,,
\end{align}
so they are conjugate to $x^{\ib}$, where $\ib=+,-,1$, as they should. 

Note that we can now identify the generators in \eqref{eq:kinetic-AdS-1} with \eqref{eq:undeformed}, and \eqref{eq:Hamiltonian}. In particular, $J^{+1}$ can be  identified with the Galilean boost $G^1$, and $\tilde{D}=D+J^{+-}$. 
%\dnote{$\mso(3,2)$ should have 10 generators. It is the conformal extension of 3-dimensional Poincar\'{e}. Conformal extension of $\mathbb{M}_4$ has 15 generators indeed.}
It can be checked that the generators \eqref{eq:Sch-gen} satisfy the following non-vanishing commutation relations
\small
\begin{subequations}\label{eq:commutation-relations-Schrodinger}
    \begin{align}
        [\tilde{D},P^+]&=-(2-\tz) P^+\,,\qquad [\tilde{D},P^-]=-\tz P^-\,,\qquad [\tilde{D},P^1]=-P^1\,,\\
        [\tilde{D},G^1]&=(\tz-1)G^1\,,\qquad \ \ \, [\tilde{D},K^+]=\tz K^+\,,\quad \ \  [P^-,K^+]=\tilde{D}+(2-\tz)\big(\beta\p_{\beta}+1\big)\,,\\
        [G^1,P^1]&=P^+\,,\qquad \qquad \quad  \ \, [G^1,P^-]=-P^1\,.
    \end{align}
\end{subequations}
\normalsize
Observe that the above Schr\"odinger algebra only closes at $\tz=2$. Away from $\tz=2$, the above algebra can close if we drop the special conformal boost generator $K$, cf. \cite{Son:2008ye}, reducing $\Sch_{\tz\neq 2}(1)$ to a smaller subalgebra where $P^{+}$ no longer corresponds to the central charge.

Then, following the usual gauging procedure, we can introduce background gauge fields and gauge parameters associated with the infinitesimal transformations for each of the above generators. For convenience, we tabulate these data in the following table:
\begin{table}[ht!]
    \centering
    \begin{tabular}{|c|c|c|c|c|}\hline
       Symmetry  & Generator & Type &Gauge field & Parameter \\ \hline\hline
       Hamiltonian  & $P_+$ & $\boldsymbol{\sD}$ & $e^{+}_{\mu}$ & $\xi^+$ \\\hline
       Translation  & $P_{1}$ & $\boldsymbol{\sK}$ & $e^{1}_{\mu}$ & $\xi^{1}$ \\\hline
        Number operator  & $P_-$ & $\boldsymbol{\sK}$ & $e^-_{\mu}$ & $\xi^-$\\\hline
       Dilatation  & $\tilde D$ & $\boldsymbol{\sK}$ & $\sigma_{\mu}$ & $\varsigma$ \\\hline 
       Galilean boost  & $G_1$& $\boldsymbol{\sK}$ & $b_{\mu}$ & $\chi$ \\\hline
       Special conformal boost  & $K_-$& $\boldsymbol{\sK}$ & $k_{\mu}$ & $\kappa$ \\\hline
    \end{tabular}
    %\caption{Caption}
\end{table}
%\anote{where the generators $P$ and $D$ are defined on the tangent space as $P_{A}=\eta_{AB}e^B{}_{\mu}P^{\mu}, G_{A}=\eta_{AB}e^B{}_{\mu}G^{\mu}$, For instance
%$$P_-=-\frac{Lv^2}{2r^4}P^++\frac{L}{r^2}P^-, P_+=\frac{L}{r^{2-z}}P^+=LP^+$$$$G_1=\frac{L}{r}G^1,\,P_1=\frac{L}{r}P^1$$}
%[\textcolor{red}{They are related to the ones in spacetime by $G_{\ib}=e_{\ib \mu}G^{\mu}=\eta_{\ib \jb}e^{\jb}{}_{\mu}G^{\mu}$. Note that we have suppressed the spacetime index $\mu$ in \eqref{eq:Adefn}.}]

Generators in the above table are defined on the \emph{tangent space}, and they obey the same algebra \eqref{eq:commutation-relations-Schrodinger}, cf. \cite{Bergshoeff:2014uea}. (We slightly abuse the notation by using the same symbols.)

For convenience, all of the above gauge fields and generators can be economically packaged in a master gauge field
%\begin{align}
%\label{eq:Adefn}
%    \mathcal{A}_{\mu}=P_{+}\tau_{\mu}+ P_1e^1{}_{\mu}+G_1\omega^{1}{}_{\mu}+P_{-}m_{\mu}+\tilde{D}b_{\mu}+Kf_{\mu}\ .
%\end{align}
%\tnote{I am getting annoyed by the fact that you keep changing the notations}
\begin{align}
\label{eq:Adefn}
    \boldsymbol{\cA}_{\mu}=e^{+}{}_{\mu}P_{+}+e^-{}_{\mu}P_-+e^1{}_{\mu}P_1+\sigma_{\mu} \tilde D+b^{1}{}_{\mu}G_{1}%+b^+G^-+c^1C^1
    +k_{\mu}\,K_-\,,
\end{align}
whose gauge transformation is:
\begin{align}\label{eq:transformation}
    \delta \boldsymbol{\cA}=d\boldsymbol{\xi}+[\boldsymbol{\cA},\boldsymbol{\xi}]
\end{align}
where
\begin{align}
    \boldsymbol{\xi}&=\xi^-P_{-}+\varsigma\tilde D+\chi G_1+\kappa K_-\,.\notag
\end{align}
These transformations are referred to as internal symmetries \cite{Bergshoeff:2014uea}. Note that the local $P^-$ and $P^1$-symmetries are not gauge symmetries, but rather time and space diffeomorphisms and therefore we will not gauge them. 
The transformation for individual fields can be read off using the commutation relations \eqref{eq:commutation-relations-Schrodinger} presented above. For instance,
\begin{subequations}
\begin{align}\label{eq:transformation-TTNC}
    \delta e^+{}_{\mu}&=\tz\,\varsigma  e^+{}_{\mu}\,,\\
    \delta e^1{}_{\mu}&=\chi e^+{}_{\mu}+\varsigma e^1{}_{\mu}\,,
\end{align}
\end{subequations}
\text{etc.} Other transformations can be obtained analogously, cf. \cite{Bergshoeff:2014uea}.

Let us now address the subtlety in gauging the Schr\"odinger symmetries \eqref{eq:Sch-gen}. Namely, when drifting away from a relativistic setting to a non-relativistic one, there is a loss in covariance. This, in turn, influences the way we should interpret the background gauge fields. First, despite their vector-like indices, the gauge fields $e^+_{\mu}$ and $e^1_{\mu}$ are in fact scalars. Second, due to the lack of spatial rotations $P^1$ and $G^1$ also behave like scalars. As a result, $(P^-,P^+,P^1,G^1,\tilde D,K)$ can all be viewed as scalars when the spatial dimension is 1. This means they can be naturally lifted to the tangent space, explaining why \eqref{eq:commutation-relations-Schrodinger} is also the algebra on the tangent space even though the original generators \eqref{eq:undeformed} and \eqref{eq:Hamiltonian} are spacetime generators. 

The above discussion may also be applicable for higher dimensional deformed $AdS$ when the spatial dimension is $d\geq 2$. In particular, while $P^-,P^+,K^+$ are again scalars in non-relativistic setting, the spacetime generators $P^i$ and $G^i$ will now transform as vectors under spatial rotations. As a result, we can lift them to the tangent space via the usual \emph{spatial} mapping \cite{Metsaev:1999ui}
\begin{align}
    T^{a_1\ldots a_n}=e^{a_1}{}_{i_1}\ldots e^{a_n}{}_{i_n}T^{i_1\ldots i_n}\,,\quad \partial_i=E_i{}^{a}\partial_{a}\,,
\end{align}
which are manifestly spatially covariant. Here, $T$ denotes some generic tensor, and we use $i,j=1,\ldots, d$ as spatial indices in spacetime while $a,b=1,\ldots,d$ as spatial indices in the tangent space. Note that $E_a{}^ie^{a}{}_{j}=\delta^i{}_j$ and $E_a{}^iE_b{}^jg_{ij}=\eta_{ab}$. One can then check that the Killing vectors in the spatial directions $\xi^a\partial_a=\xi^{\mu}\partial_{\mu}$. Thus, generators on the tangent space in the spatial directions have the same form with the ones in spacetime. This allows one to gauge \eqref{eq:commutation-relations-Schrodinger} on the tangent space. 

%%%%%%%%%%%%%%%%%%%%%%%%%%%%%%%%%%%%%%%%%%%%%%
\paragraph{Twist-free torsional Schr\"odinger geometry.} To make the above geometry become that of Newton-Cartan, cf. \cite{Cartan:1923zea}, we can start to impose appropriate constraints on the curvature two-form
\begin{align}\label{eq:Cartan}
    \boldsymbol{\cR}=d\boldsymbol{\cA}+ \boldsymbol{\cA}\wedge\boldsymbol{\cA}\,.
\end{align}
Due to the presence of the $(dx^+)^2$ term in the metric \eqref{eq:AdS-metric-2} and the Lifshitz scalings \eqref{eq:M-scale}, the Schr\"odinger geometry that we are about to study will intrinsically have torsion, regardless of what conditions we will impose on \eqref{eq:Cartan}. 
The minimal set of constraints that lead to a reasonable torsional Newton-Cartan geometry turns out to be 
\begin{align}\label{eq:flat-R}
    \boldsymbol{\cR}[P^+]=0\,,\qquad \boldsymbol{\cR}[P^1]=0\,.
\end{align}
As the vierbein are the only local fields that transform under $P$-transformation, cf. \eqref{eq:transformation}, we can treat them as independent fields. Then, we can solve (at $\tz=2$) the set of constraints \eqref{eq:flat-R} for various components of the corresponding gauge fields $\boldsymbol{\cA}$ through
\begin{subequations}\label{eq:to-solve-twist-free}
    \begin{align}
        P^+&: &\cR^-{}_{\mu\nu}&=\p_{[\mu}e^-{}_{\nu]}%-(2-\tz) \sigma_{[\mu}e^-_{\nu]}
        +
        b_{[\mu}e^1{}_{\nu]}=0\,,\\%+b^{+}_{[\mu}e^-_{\nu]}\,,\\%+c_{[\mu}^{1}e^1_{\nu]}-b^{+}_{[\mu}e^+_{\nu]}\,,\\
        P^1&: &\cR^1{}_{\mu\nu}&=\p_{[\mu}e^1{}_{\nu]}- \sigma_{[\mu}e^1{}_{\nu]}%-b_{[\mu}^{+}e^+_{\nu]}
        -b_{[\mu}e^+{}_{\nu]}=0\,.
    \end{align}
\end{subequations}
%As a result, the geometry is not Newton-Cartan \`a la \cite{Bergshoeff:2014uea}. 
where $[\,,]$ denotes the anti-symmetric combination \emph{without} a factor of $\frac{1}{2}$. Plugging \eqref{eq:vierbein} into the above, we can determine from \eqref{eq:to-solve-twist-free} that 
\begin{subequations}\label{eq:twist-free-solution}
    \begin{align}
        \sigma_{\mu}&=-\frac{\delta^r{}_{\mu}}{r}\,,\label{eq:dilaton}\\
        b_{\mu}&=0\,.
    \end{align}
\end{subequations}
Note that the bulk dilaton $\sigma$ can be associated with the 't Hooft coupling $\lambda$ on the boundary theory, which is dual to the gravitational theory that we aim to construct. 
Furthermore, our solution is different from \cite{Bergshoeff:2014uea} in the sense that the dilaton can be solved explicitly, and we do not impose $\boldsymbol{\cR}[P^-]=0$ as well as $\boldsymbol{\cR}[\tilde D]=0$ from the outset.\footnote{Due to the form of $e^+_{\mu}$, one cannot solve for the dilaton uniquely if $\boldsymbol{\cR}[P^-]=0$ is also imposed. Moreover, with the provided solution, it can be checked that $\boldsymbol{\cR}[\tilde D]=0$ without specifying the field $k_{\mu}$. For convenience, we will also set $k_{\mu}=0$.} Thus, naturally, we expect the background will be intrinsically torsional. With the solution \eqref{eq:twist-free-solution}, we have
\begin{align}\label{eq:vierbein-postulate}
    \cR^+{}_{\mu\nu}[P^-]-T^+{}_{\mu\nu}&=\p_{[\mu}e^+{}_{\nu]}-2%\tz 
        \sigma_{[\mu}e^+{}_{\nu]}-\upsilon^2L\frac{\sigma_{[\mu}\delta^+{}_{\nu]}}{r^4}=0\,,
\end{align}
which may be viewed as a vierbein postulate for $e^+{}_{\mu}$. Here, we may interpret
\begin{align}
    T^{\rho}{}_{\mu\nu}:=\frac{\upsilon^2L}{r^4}\sigma_{[\mu}\delta^{\rho}{}_{\nu]}\,, 
\end{align}
as a torsion two-form, which arose due to the Lifshitz anisotropic scalings \eqref{eq:M-scale}. Note that our way of defining the torsion 2-form is different with \cite{Bergshoeff:2014uea} in the sense that it comes out from the flatness completion of $\boldsymbol{\cR}[P^-]$. Observe that our \emph{``clock''} $e^+_{\mu}$ is not $d$-closed. Nevertheless, it is integrable since $e^+\wedge de^+=0$. As a result, the Schr\"odinger geometry is said to be \emph{twist-free}. The physical implications of the above twist-free torsional geometry will be explained in Section \ref{sec:holo}.

%%%%%%%%%%%%%%%%%%%%%%%%%%%%%%%%%%%%%%%%%%%%%%%%%
\section{Non-relativistic dynamics on the Schr\"odinger cone}\label{sec:NR-LC}

The presence of torsion in the geometry calls for a revisiting of the analysis in \cite{Metsaev:1999ui,Metsaev:2008fs,Metsaev:2018xip} to identify the crucial modifications in the dynamics relevant for constructing theories in the light-cone gauge, or rather the Schr\"odinger cone. (See also \cite{Hagen:1972pd} for previous work on free higher-spin fields on torsional Newton-Cartan background, as well as the works \cite{Gutperle:2013oxa,Beccaria:2015iwa}.)

%%%%%%%%%%%%%%%%%%%%%%%%%%%%%%%%%%%%%%%%%%%%%%%%%%%%%
\subsection{Kinetic action in Schr\"odinger background}

From the dynamical generator $P^-$, it is expected that the Lagrangian for any spinning field on the Schr\"odinger background \eqref{eq:AdS-metric-2} should be of the form 
\begin{align}\label{eq:Lagrangian-s}
    \cL[s]=\varphi_{-s}\Big(\Box+\upsilon^2 r^{2-2\tz}(\p_-)^2\Big)\varphi_{+s}\,,\qquad \Box=2\p_+\p_-+\p_1^2+\p_r^2\,.
\end{align}
where $\varphi_{+s}$ and $\varphi_{-s}$ are two complex valued fields that are Hermitian conjugate to each other. Notice that the kinetic operator $\Box+\upsilon^2r^{2-2\tz}(\p_-)^2$ has dimension $[length]^{-2}$ at $\tz=2$. 
To show that the above is a valid Lagrangian in the light-cone gauge, let us study a simple model of a conformally coupled scalar field $\varphi$ on the Schr\"{o}dinger background \eqref{eq:AdS-metric-2} as an example by looking at the following action
\begin{align}
\label{eq:originalphiaction}
    S_2[0]=\int d^4x\sqrt{-g}\Big(g^{\mu\nu}\p_{\mu}{\varphi}\p_{\nu}{\varphi}+\frac{1}{6}R{\varphi}^2\Big)\,.
\end{align}
For the background \eqref{eq:AdS-metric-2}, the Ricci scalar $R=-\frac{12}{L^2}$. Then, upon rescaling $\varphi\mapsto \frac{r}{L} \varphi$, the above action reduces to 
\begin{equation}
\begin{aligned}\label{eq:free-scalar-NR}
    S_2[0]&=-\frac{1}{2}\int d^4x\,\varphi\Big(\Box+\upsilon^2r^{2-2\tz}(\p_{-})^2\Big) \varphi \,.
\end{aligned}
\end{equation}
%where $\Box\equiv 2{\partial_{+}}{\partial_{-}}+\partial_1^2+\partial_r^2$.
It is important to point out that $\sqrt{-g}$ has been absorbed by the above rescaling and there is no AdS radius $L$ remaining. Observe that without the $(\partial_-)^2$ term, the kinetic operator is simply the one of flat space.\footnote{Note the positions of the indices for such simplification to happen.}

At this stage, we can take \eqref{eq:Lagrangian-s} to be the kinetic action for any spinning fields in the light-cone gauge, since we have already derived the explicit form of the $P^-$ generator; keeping in mind that any spinning fields can be described by two scalars, which are complex conjugate to each other. (Indeed, while the $M^{r1}$ generator, cf. \eqref{eq:helicity-operator}, does not participate in the Schr\"odinger algebra, it nevertheless remains present as the operator that specifies the helicity of fields.) Then, as usual, if we start with a generic tensorial field and imposing the light-cone gauge, one can choose two components among the remaining components of the original tensor field to be the physical ones -- the $\varphi_{\pm s}$. For relevant discussion, see e.g. Section 13 in \cite{Ponomarev:2022vjb}.

In making connection with the covariant approach, the reader may find the material in Appendix \ref{app:B} helpful.\footnote{There, we show that after some suitable rescaling, all chiral spinning fields (specific field representations that are closely related to our setting) can have a similar kinetic term with the ones in flat space. This aligns with the result in \cite{Metsaev:1999ui}. However, due to the lack of Lorentz and gauge invariance, we could not engineer some terms which give the differential operator $\upsilon^2r^{2-2\tz}(\partial_-)^2$. }
%%%%%%%%%%%%%%%%%%%%%%%%%%%%%%%%%%%%%%%%%%%%%%%%%
\subsection{Bulk-to-boundary propagator and Schr\"odinger boundary algebra}
%\anote{What is the definition of spin here when we do not have rotation generator} \tnote{Spin is $M^{r1}$ see footnote 2 above.}
%\dnote{But we have $M^{r1}$ for a relativistic system. Once deformed into Sch, generators are (3.8). Now, these have no explicit $M^{r1}$. So, how should we think of spin in case of non-relativistic system with no effective rotation generators/helicity?} \tnote{I think the helicity label just follow to NR case. See \cite{Hagen:1972pd}. Also, in LC they are all scalars. On the CFT side, there are higher-spin currents \cite{Bekaert:2011qd}, so there should be spinning bulk fields.}

Let us now compute the bulk-to-boundary propagator for the \emph{rescaled} higher-spin fields in the light-cone gauge, whose Lagrangian is \eqref{eq:Lagrangian-s}. 

%In this section we elaborate on the computation of the 2-point function of the operator dual to the \emph{rescaled} scalars $\varphi$. 
To compute the bulk-to-boundary propagator for $\varphi_{\pm s}$, we will follow the techniques developed in \cite{Fuertes:2009ex,Volovich:2009yh,Leigh:2009eb}. Note that since we are working with \emph{rescaled} complex scalars, there are some intermediate details that differ from the standard treatment in general dimensions, in e.g. \cite{Fuertes:2009ex,Volovich:2009yh,Leigh:2009eb}. Recall that the equation of motion of a rescaled scalar $\varphi_{\pm s}$ is given by
\begin{equation}
\label{eq:varphi-EOM}
    \Big(\Box+\upsilon^2r^{2-2\tz}(\p_{-})^2\Big) \varphi_{\pm s} =0\,,\qquad \text{where} \qquad \Box \equiv 2\partial_{+}\partial_{-}+\partial_1^2+\partial_r^2\, .
\end{equation}
Since there are translation invariances away from the radial direction, cf. \eqref{eq:AdS-metric-2}, the solution can be decomposed into plane wave modes along boundary directions. In particular,
\begin{equation}\label{eq:phiansatz}
    \varphi(x^+,x^{-},x^1,r) =\int \frac{d^3p}{(2\pi)^{3/2}}~  e^{i(\gamma x^{+}+\beta x^{-}+\rho x^1)}\cK_p(r)\,,\qquad \vec{p}=(\beta,\gamma,\rho)\,.
\end{equation}
where we have denoted $ \gamma\equiv p^-$. The radial mode $\cK_p(r)$ of a complex scalar field thus satisfies the following equation
\begin{align}
\big(2\beta\gamma+\rho^2+r^{2-2\tz}\upsilon^2\beta^2-\p_r^2\big)\cK_p(r)=0\,.
\end{align}
For generic integer values of the dynamical exponent $\tz$, solutions to the above exist but are quite complicated. Here, we are mainly interested in the case where $\tz=2$, for which the solution is given by,
\begin{equation}
    \cK_p(r)= c_1(\vec{p}) \sqrt{r}I_{\nu}(|p|r) + c_2(\vec{p})\sqrt{r}K_{\nu}(|p|r)\ ,
\end{equation}
where
\begin{equation}
    \nu=\frac{1}{2}\sqrt{1+4\upsilon^2\beta^2}\,, \quad \text{and}\quad |p|\equiv \sqrt{p^2}=\sqrt{2\beta\gamma+\rho^2}\ .
\end{equation}
Here, $I_{\nu}$ and $K_{\nu}$ are \emph{modified Bessel functions} of the first and second kind, respectively. 

Before proceeding further, it must be noted that our deformed $AdS_4$ spacetime has Lorentzian signature, which brings with it certain subtleties as compared to Euclidean case. In particular, regularity of the normalizable solutions in the deep bulk region ($r \to \infty$) demands that $c_1(\vec{p})=0$.\footnote{This follows from the asymptotic expansions for large $r$ of the modified Bessel functions: 
    \begin{align}
        I_{\nu}(|p|r) \sim \frac{e^{|p|r}}{\sqrt{|p|r}}\,,\qquad K_{\nu}(|p|r) \sim \frac{e^{-|p|r}}{\sqrt{|p|r}}\,.
    \end{align}} 
    As a result, to make the solution regular, we shall impose another boundary condition, which is $\cK_p|_{r=\epsilon}=1$ at a cutoff $r=\epsilon$. This leads to a complete solution, which is
\begin{equation}
\label{eq:bulk-to-bdy-prop}
    \cK_p(r)=\frac{\sqrt{r}K_{\nu}(|p|r)}{\sqrt{\epsilon}K_{\nu}(|p|\epsilon)}\ .
\end{equation}
Note that the cutoff $\epsilon$ is important in computing the two-point function \cite{Klebanov:1999tb}, or the loop integrals as to regularize divergences. However, for other higher-point tree-level amplitudes in AdS, we can safely absorb the factor $\sqrt{\epsilon}K_{\nu}(|p|\epsilon)$ in the definition of the higher-point functions. Furthermore, timelike ($p^2<0$) or spacelike ($p^2>0$) 3-momentum along the boundary directions will not affect our analysis below too much since the solution \eqref{eq:bulk-to-bdy-prop} has been amended appropriately such that it is regular everywhere. To this end, the general solution to the equation of motion \eqref{eq:varphi-EOM} is
\begin{equation}
\begin{aligned}
    \varphi(x^+, x^{-}, x^1, r)%\\
    %&\hspace*{0.2cm}
    =\int \frac{d^3p}{(2\pi)^{3/2}}~ e^{i(\gamma x^+ + \beta x^{-}+\rho x^1)}\frac{\sqrt{r}K_{\nu}(|p|r)}{\sqrt{\epsilon}K_{\nu}(|p|\epsilon)}\,.%\left[c_2(\vec{p})K_{\nu}(|p|r)+ \theta(-p^2)c_1(\vec{p})I_{\nu}(i|p|r)\right]
\end{aligned}
\end{equation}

%\dnote{Things to write: contour deformation for Lorentizian, contrasting with Euclidean, absence of normalizable mode, final time ordered correlator.} \tnote{I fixed some statement a little bit.}

Next, we will find the conformal dimension of the operator dual to the bulk \emph{rescaled} complex scalars $\varphi_{\pm s}$ by examining the behavior of $\varphi_{\pm s}$ near the boundary $r=0$. Assuming the solution scales as $\varphi=r^{\Delta}{\phi}_0(x^+,x^-,r)$ near the boundary, we obtain
\begin{align}
    \Delta(\Delta-1)-\upsilon^2\beta^2=0\,.
\end{align}
The roots of the above are:
\begin{align}\label{eq:Delta-roots}
    \Delta_{\pm}=\frac{1\pm \sqrt{1+4\upsilon^2\beta^2}}{2}\,.
\end{align}
For simplicity, we may set $\upsilon=1$, assuming that we are working at $\tz=2$ and $\beta^2$ is a number. (In fact, after a dimensional compactification along the $x^-$ direction, $\beta$ can be replaced with a dimensionless parameter viewed as the non-relativistic mass $m_s$ associated with $\varphi_{\pm s}$, cf. \cite{Son:2008ye}.) Observe that due to the anisotropic scaling, the above roots do not satisfy the usual relation $\Delta(\Delta-d)=m^2L^2$, which determines the conformal dimension of a massive scalar field in $AdS_{d+1}$. 

In the same spirit with Flato-Fronsdal theorem \cite{Flato:1978qz}, keeping in mind the fact that we are in an IR regime after the deformation of geometry happened, cf. \eqref{eq:AdS-metric-2}, we will choose $\Delta_0=\Delta_++1$ to be the conformal dimension of the rescaled scalar field, and $\Delta_s=\Delta_{-}+1$ to be the conformal dimension of the higher-spin fields in what we call the Metsaev frame.\footnote{This is the frame where the kinetic term for higher-spin fields in $AdS$ reduces to that of flat space, and all higher-spin fields have the same conformal dimensions upon rescaling \cite{Metsaev:1999ui}. Recall the manipulation around \eqref{eq:originalphiaction}.} This requires us to modify the bulk-to-boundary propagator for spinning fields as \cite{Skvortsov:2018uru}
\begin{align}
    \cK_p^{\pm s}=\frac{1}{p}\frac{\sqrt{r}K_{\nu}(|p|r)}{\sqrt{\epsilon}K_{\nu}(|p|\epsilon)}\,.
\end{align}
to match the boundary two-point function normalization. 
Here, the rescaled higher-spin fields in the bulk correspond to some rescaled spin-$s$ operators $\cJ_{s\geq 1}$ on the boundary side in the transverse-traceless basis, see \cite{Metsaev:1999ui}. Note also that unlike the case of exact $AdS$, here, the dual operators on the field theory side can be irrelevant depending on the value of mass that is reasonably assumed to be positive for NR theories. 
To this end, let us drag the propagator through the generators of the Schr\"odinger algebra 
\begin{align}
    \cG_{AdS} \Big(\varphi=\cK\,\cJ(p)\Big)=\cK \,\cG_{b}\,\cJ(p)\,,\qquad \vec{p}=(\beta,\gamma,\rho)\,.
\end{align}
We find that 
\begin{subequations}
    \begin{align}
        P^+_b&=\beta\,,\qquad \qquad P^-_b=\gamma\,,\qquad \qquad P^1_b=\rho \,,\\
        G^1_b&=\beta\p_{\rho}-\rho\p_{\gamma}\,,\\
       \tilde D_b&=\Big(-(2-\tz)\beta\p_{\beta}-\rho\p_{\rho}-\tz \gamma\p_{\gamma}-\frac{1}{2}-\tz+\nu\Big)\Big|_{\tz=2}\,,\\
        K_b^+&=-\frac{1}{2}\beta\Big(2(2-\tz)\p_{\gamma}\p_{\beta}+\p_{\rho}^2\Big)-\tilde D_b\p_{\gamma}\,,
    \end{align}
\end{subequations}
are the generators on the boundary ($b$ for boundary). It can be checked that
\small
\begin{subequations}
    \begin{align}
        [\tilde D_b,P^+_b]&=-(2-\tz) P^+_b\,,\qquad \ [\tilde D_b,P^-_b]=-\tz\, P^-_b\,,\qquad [\tilde D_b,P^1_b]=-P^1_b\,,\\
        [\tilde D_b,G^1_b]&=(\tz-1)G^1_b\,,\quad \qquad \,  [\tilde D_b,K_b^+]=\tz\, K_b^+\,,\qquad
        [P^-_b,K^+_b]=\tilde{D}_b+J^{+-}_b\,,\\
        [G^1_b,P^1_b]&=P^+_b\,,\qquad \qquad \quad  \ \, \ \  [G^1_b,P^-_b]=-P^1_b\,,
    \end{align}
\end{subequations}
\normalsize
matching with the Schr\"odinger algebra $\Sch_{\tz}(1)$ on the bulk side at $\tz=2$. Note that the generator $J_b^{+-}=-\tz \gamma\p_{\gamma}+(2-\tz)\beta\p_{\beta}$ is the CFT dual of $J^{+-}$ in \eqref{eq:Jpm} in the Schr\"odinger setting.
%%%%%%%%%%%%%%%%%%%%%%%%%%%%%%%%%%%%%%%%%%%%%%%%%
\subsection{Cubic vertices in Schr\"odinger background}

In the light-front approach for field theories with local interactions, the dynamical generators will acquire corrections that are higher-order in fields, see \cite{Metsaev:2005ar,Ponomarev:2022vjb} for a review. In our non-relativistic setting, the interactions encoded in higher-order Hamiltonians $P^-_{n\geq 3}$ are required to satisfy \cite{Metsaev:2018xip}
\begin{align}
   [\boldsymbol{\sK},\boldsymbol{\sD}]=\boldsymbol{\sD}\quad \Rightarrow \quad  \begin{cases} [\boldsymbol{\sK},\boldsymbol{P}^-]=0\,,\qquad \qquad  \boldsymbol{\sK}=\{G^1,K^+\}\,,\\
    [\tilde D,\boldsymbol{P}^-]=-2\boldsymbol{P}^-\,,\label{eq:dilation-constraint}
    \end{cases}
\end{align}
%\anote{$\boldsymbol{\sK}=\{G^1,K^+\}$?}
where $[\,,]$ denotes the Poisson-Dirac bracket, cf. \eqref{eq:Poisson-Dirac}. Here,
\begin{align}
     \boldsymbol{P}^-=P_2^-+\sum_{n\geq 3}P^-_n\,,
\end{align}
where $P_2^-$ is the Hamiltonian generator that is quadratic in the fields. In the above, the constraint $[\tilde D,\boldsymbol{P}^-]=-2\boldsymbol{P}^-$ is a bit special and deserved a clarification. In position space, a generic action at order $n$ in fields can be written as $S_n=\int dx^+ dx^-dx^1dr~ V_n(\varphi_1,\ldots,\varphi_n)$. A simple counting tells us that the associated Hamiltonian,
\begin{align}
    P_n^-=\int dx^-dx^1dr V_n(\varphi,\ldots,\varphi)
\end{align}
will have mass dimension two. 

Next, as all dynamical boost generators are absent once the system is driven away from the exact $AdS_4$ background, we will not have enough constraints to fix the cubic couplings uniquely as in \cite{Metsaev:2018xip}. Nevertheless, the leading contribution should follow the analysis of \cite{Metsaev:2018xip}, with some small modification due to the Lifshitz anisotropic scalings. 
In momentum space, we shall denote
\begin{align}
    P^-_n&=\int d\Gamma_n\,\Tr\Big(\prod_{a=1}^n\varphi_a^{\dagger}\Big)p^-_n(\tp_a,\p_{r_a};r)\delta_r\,,\qquad \delta_r:= \prod_{a=1}^n\delta(r-r_a)\,,
\end{align}
where $\vec{p}_a=(\beta_a,\gamma_a,\rho_a)$, and
\begin{align}
    d\Gamma_n=(2\pi)^2\delta^2\Big(\sum_{a=1}^n \tp_a\Big)\prod_{a=1}^n\frac{d^2\tp_a}{2\pi}dr_a\,dr\,,\qquad d^2\tp_a=d\rho_a d\beta_a\,.
\end{align}
Note that $a$ is the label of the external fields.

%%%%%%%%%%%%%%%%%%%%%%%%%%%%%%%%%%%%%%%%%%%%%%%%%%
Next, from $[\boldsymbol{\sK},\boldsymbol{\sD}]=\boldsymbol{\sD}$, we get the following constraints for higher-order Hamiltonians (at $\tz=2$)
\begin{subequations}
    \begin{align}
  \sum_{a=1}^n \beta_a\p_{\rho_a}p^-_n(\tp_a,\p_{r_a};r)\delta_r&=0\,,\label{eq:J+1}\\
  \sum_{a=1}^n \Big(-\rho_a\p_{\rho_a}+\p_{r_a}r_a\Big)p^-_n(\tp_a,\p_{r_a};r)\delta_r&=\big(n-3\big)p^-_n(\tp_a,\p_{r_a};r)\delta_r\,,
   \label{eq:D}\\
    \sum_{a=1}^n\,\beta_a\big(\p^2_{\rho_a}-r_a^2\big)p^-_n(\tp_a,\p_{r_a};r)\delta_r&=0\,,\label{eq:K}
\end{align}
\end{subequations}
\normalsize
using \eqref{eq:Poisson-Dirac} and doing integration by part. In particular, we have used the simple relations
\begin{align}
    [F(\varphi),\boldsymbol{\sK}]=[\varphi,\boldsymbol{\sK}]\frac{\delta F(\varphi)}{\delta \varphi}=\boldsymbol{\sK}\varphi\frac{\delta F(\varphi)}{\delta \varphi}\,,
\end{align}
where $F(\varphi)$ is some generic function in $\varphi$, and $[\,,]$ is the Poisson-Dirac bracket \eqref{eq:Poisson-Dirac}. Note that as $d\Gamma_n$ contains $\delta^2\big(\sum_{a=1}^n\tp_a\big)$, there is an extra $-1$ when $\tilde D$ acts on it. Recall that $\beta$ is dimensionless, so $\tilde D$ only counts the dimension of $\delta(\sum_a\rho_a)$. 

In what follows, we will only focus on the case $n=3$. Equation \eqref{eq:J+1} tells us that the density $p^-_3(\tp_a,\partial_{r_a};r)\delta_r$ will be polynomial in 
\begin{align}
    \PP=\frac{1}{3}\big(\check\beta_1 \rho_1+\check\beta_2\rho_2+\check\beta_3\rho_3\big)\,,\qquad \check\beta_a=\beta_{a+1}-\beta_{a+2}\,, \quad a=1,2,3\,.
\end{align}
With this in mind, we can replace $\sum_{a=1,2,3}\rho_a\p_{\rho_a}\mapsto \PP\p_{\PP}$ and $\sum_{a=1,2,3}\beta_a\p_{\beta_a}\mapsto \PP\p_{\PP}$ by a mere change of variables. Then, at $\tz=2$, \eqref{eq:D} becomes
\begin{align}
   &\Big( -(3-\tz)\PP\p_{\PP}+\sum_a\p_{r_a}r_a\Big)p^-_3(\tp_a,\p_{r_a};r)\delta_r\Big|_{\tz=2}=0\,.
\end{align}
\normalsize
This indicates that the Lifshitz deformation under consideration can preserve some part of the cubic vertices in exact $AdS_4$, cf. \cite{Metsaev:2018xip}. In particular, we get
\begin{align}
    N_{\PP}=N_r+1\,,\qquad N_{\PP}=\PP\frac{\partial}{\p \PP}\,,\quad N_r=r\frac{\p}{\p r}\,.
\end{align}
Next, using momentum conservation, equation \eqref{eq:K} gives us
\begin{align}\label{eq:K-sym}
    \big(B\p_{\PP}\p_{\PP}+\sum_{a=1}^3\beta_a r_a^2\big)p^-_3(\tp_a,\p_{r_a};r)\delta_r=0\,,\qquad B=\beta_1\beta_2\beta_3\,.
\end{align}
One can then show that the sum of quadratic Hamiltonian reduces to the usual one in $AdS_4$, i.e.
\begin{align}
    \sum_{a=1,2,3}P^-_a=\sum_{a=1,2,3}\Big(-\frac{\rho_a^2}{2\beta_a}-\frac{1}{2}r^{2-2\tz}\beta_a+\frac{\p_{r_a}^2}{\beta_a}\Big)%=\sum_a\Big(-\frac{\rho_a^2}{2\beta_a}+\frac{\p_{r_a}^2}{\beta_a}\Big)
    =\frac{\PP^2}{2B}+\sum_{a=1,2,3}\frac{\p_{r_a}^2}{\beta_a}\,.
\end{align}
\normalsize
Following \cite{Metsaev:2018xip}, one can show that the above can be written as 
\begin{align}
   \sum_{a=1,2,3}P^-_a= \frac{\PP^2-\PP_r^2}{2B}+\frac{\Delta_{\beta}}{36B}\p_r^2+\frac{1}{3}\cP_r\p_r\,,
\end{align}
where
\begin{align}
    \PP_r=\frac{1}{3}\sum_a\check\beta_a\p_{r_a}\,,\quad \cP_r=\sum_a\frac{\p_{r_a}}{\beta_a}\,,\qquad \Delta_{\beta}=\beta_1^2+\beta_2^2+\beta_3^2\,,\quad B=\beta_1\beta_2\beta_3\,.
\end{align}
This indicates that $p^-_3(\tp_a,\p_{r_a};r)=p^-_3(\PP,\PP_r,\cP_r,\p_{r_a},r_a)$. Assuming $p^-_3(\PP,\PP_r,\cP_r,\p_{r_a},r_a)$ is harmonic in $(\PP,\PP_r)$, i.e. 
\begin{align}
    (\p_{\PP}^2-\p_{\PP_r}^2)p_3^-=\big(\p_{\PP}-\p_{\PP_r}\big)\big(\p_{\PP}+\p_{\PP_r}\big)p^-_3=2\p_{\PP^R}\p_{\PP^L}\,p^-_3(\PP,\PP_r,\cP_r,\p_{r_a},r_a)=0\,,
\end{align}
it can then be shown, using \eqref{eq:K-sym}, that $p^-_3(\PP,\PP_r,\cP_r,\p_{r_a},r_a)$ is independent of $\cP_r$, cf. \cite{Metsaev:2018xip}, and 
\begin{align}
    %p^-_3=p^-_3(\PP^L,\PP^R,\beta_a,r)\,,\qquad 
    \Big(N_r-N_{\PP^L}-N_{\PP^R}+1\Big)p^-_3(\PP^L,\PP^R,\beta_a,r)=0\,,
\end{align}
where 
\begin{align}
    (\PP^L,\PP^R)=\frac{1}{\sqrt{2}}\Big(\PP+\PP_r,\PP-\PP_r\Big)\,,\qquad N_r=r\p_r\,,\quad N_{\PP^{\bullet}}=\PP^{\bullet}\p_{\PP^{\bullet}}\,,\quad \bullet=L,R\,.
\end{align}
At this stage, one has the option to split the cubic vertex into a holomorphic piece, denoted as $V(\PP^L,\beta_a,r)$, and an anti-holomorphic piece, denoted as $\bar{V}(\PP^R,\beta_a,r)$. These vertices obey the following constraints
\begin{align}\label{eq:NP-constraint}
    \Big(N_r-N_{\PP^L}+1\Big)V=0\,,\qquad \Big(N_r-N_{\PP^R}+1\Big)\bar{V}=0\,.
\end{align}
Thus, in a Lifshitz anisotropic setting, the relations between the power of $(\PP^L,\PP^R)$ and the power of $r$ in the cubic vertices remains the same as in the case of exact $AdS_4$ \cite{Metsaev:2018xip}. To this end, we note that we do not have enough dynamical boost generators to completely fix the cubic interactions. This brings us to the next point.

%%%%%%%%%%%%%%%%%%%%%%%%%%%%%%%%%%%%%%%%%%%%%%%%%
\section{A non-relativistic chiral massive higher-spin theory}\label{sec:theory}

As shown in Section \ref{sec:3}, the twist-free torsional Schr\"odinger geometry is geometrized by the vierbein \eqref{eq:vierbein}, and the dilaton \eqref{eq:dilaton}. These dynamical fields naturally %induce non-relativistic scaling symmetries and 
\emph{``Higgs''} masses for massless relativistic fields when they propagate on Schr\"odinger background, effectively drive the massless chiral higher-spin gravity in $AdS_4$ to a massive one. 

Since we must keep the number dof. intact, the resulting massive system should live in a 3 dimensional spacetime. As $x^-$ is \emph{frozen} at $\tz=2$ under the Lifshitz scaling following \eqref{eq:M-scale}, it is thus natural to compactify the $x^-$-direction of the Schr\"odinger cone, cf. \cite{Son:2008ye}, to bring our $4d$ system to a $3d$ spacetime. Note that this procedure differs from standard Kaluza-Klein compactification in relativistic theories in the sense that the kinetic term is first order in time derivative $\partial_+$, leading to a massive non-relativistic field theory in a Lifshitz deformed $AdS_3$, whose line element is
\begin{align}\label{eq:metric-AdS3}
    ds^2=-\frac{(dx^+)^2}{r^4}+\frac{1}{r^2}\big(dx_1^2+dr^2\big)\,.
\end{align}
Here, we have set $\upsilon=1$ and $L=1$, as there is no ambiguity.
%%%%%%%%%%%%%%%%%%%%%%%%%%%%%%%%
\paragraph{Kinetic action.} Let us elaborate the above idea. Decomposing fields into their Kaluza-Klein modes on a circle along $x^-$ direction:
\begin{align}\label{eq:KK-modes}
    \varphi_{\pm s}=\sum_{m_s}e^{\pm im_sx^-}\varphi_{\pm s}\,,\qquad [m_s]=[length]^0\,,
\end{align}
and feeding them back to the free action for relativistic fields, the resulting action becomes an action for non-relativistic fields of the form
\begin{align}\label{eq:NR-free-action}
    S_2[s]=\int \varphi_{-s}\Big(2i\,m_s\p_++ \p_1^2+\p_r^2-\frac{1}{r^2}\upsilon^2 m^2_s\Big)\varphi_{+s}\,.
\end{align}
Here, we do not exclude the possibility that the mass $m_s$ is spin dependent. Moreover, it is easy to notice that the zero modes in the discrete light-cone quantization \eqref{eq:KK-modes} is non-dynamical, and thus, can be ignored. This subtlety is rather crucial in obtaining reasonable non-relativistic physics. To this end, we shall emphasize that the mechanism that generates mass for higher-spin fields considered here is different with the one of ``topological'' massive gravity in \cite{Skenderis:2009nt}.\footnote{Despite the name, the massive graviton in \cite{Skenderis:2009nt} does, in fact, propagate.}
%%%%%%%%%%%%%%%%%%%%%%%%%%%%%%%%
\paragraph{Interactions.} 
As shown in Section \ref{sec:NR-LC}, it is complicated even with the help of the light-front approach to make explicit statements at the interacting level, as the Lifshitz exponential scaling tends to drift the original relativistic system onto a twist-free torsional Schr\"odinger background, unavoidably breaking gauge invariance and Lorentz invariance. Nonetheless, we can still borrow the $4d$ cubic vertices in \cite{Metsaev:2018xip} and perform a dimensional reduction to obtain part of the interaction vertices for a massive theory in a deformed $AdS_3$ via \eqref{eq:KK-modes}. 

As mentioned in the Introduction, there exists a unique local massless higher-spin gravity in $4d$ that admits, to some extent, a field-theoretic interpretation. This theory is chiral HSGRA \cite{Metsaev:2018xip,Skvortsov:2018uru}. At cubic level, chiral HSGRA contain only one type of vertices, be them holomorphic or anti-holomorphic. Here, we consider the holomorphic sector, with the following cubic action:
\begin{align}
    S_3=\int dr\prod_{a=1}^3dr_a\,V_3[h_1,h_2,h_3]\prod_{a=1}^3\delta(r-r_a)\,\varphi_{h_1}\varphi_{h_2}\varphi_{h_3}\,,
\end{align}
where
\begin{align}\label{eq:V_3}
    V_3[h_1,h_2,h_3]=\frac{1}{\Gamma(\tH)}\times \frac{1}{r}\sum_{n=0}^{\tH-1}\frac{(r\PP^L )^{\tH-n}}{\beta_1^{h_1}\beta_2^{h_2}\beta_3^{h_3}}\cQ^{n}\,,\qquad \tH=h_1+h_2+h_3>0\,.
\end{align}
Here, $\cQ(\cC,\check{B},\cH)$ is referred to as Metsaev polynomial, which is a certain combination of 
\begin{align}
   \cC=\frac{1}{3}\sum_{a=1,2,3}\check{\beta}_ah_a\,,\quad \check{B}=\check{\beta}_1\check{\beta}_2\check{\beta}_3\,,\quad \cH=\sum_{a=1,2,3}\frac{h_a}{\beta_a}\,,
\end{align}
reflecting the non-commutativity of covariant derivatives in $AdS$. Note that the cosmological constant or $AdS$ radius does not appear explicitly in any of the cubic vertices because we are working with the rescaled fields $\varphi_{h_a}$. However, we should remember that they have conformal dimensions $\Delta=\frac{3}{2}\pm \nu$, as mentioned in the previous section. Furthermore, only terms with total helicity $\tH > 0$ can contribute non-trivially to the cubic vertices. 

Now, to drive chiral HSGRA from exact $AdS_4$ to the deformed $AdS_3$, we can simply replace\footnote{This is slightly different with the relativistic compactification considered in \cite{Metsaev:2020gmb,Skvortsov:2020pnk}.}
\begin{align}
    \beta_a\mapsto \,\sign(h_a)m_a(s_a)\,,
\end{align}
assuming that momentum conservation is preserved even after dimensional compactification. Then, $(\PP,\PP_r)$ reduce to
\begin{align}
    \PP\mapsto \PP[m]=\frac{1}{3}\sum_{a=1,2,3}\check{m}_a\rho_a\,,\qquad \PP_r\mapsto\PP_r[m]=\frac{1}{3}\sum_{a=1,2,3}\check{m}_a\p_{r_a}\,.
\end{align}
where
\begin{align}
    \check{m}_a=\sign(h_{a+1})m_{a+1}-\sign(h_{a+2})m_{a+2}\,.
\end{align}
As the result, the holomorphic cubic vertices of our $3d$ NR massive chiral HSGRA have the form
\begin{align}\label{eq:NR-cubic}
    \cV_3[h_1,h_2,h_3]=V_3[h_1,h_2,h_3]+\text{corrections}\,,
\end{align}
where the ``correction'' part stands for the tail coming from commuting covariant derivatives in the deformed $AdS_3$, cf. \eqref{eq:metric-AdS3}. Here, we still have the notions of helicity in $3d$ thanks to the fact that our higher-spin fields are massive and have two degrees of freedom. That said, our massive gravitational theory in $AdS_3$ is not topological. 

To this end, we shall emphasize that symmetry constraints in the bottom-up approach are insufficient to make explicit statements about the correction terms without additional input. However, if one assumes that the AdS/CFT correspondence remains valid in the non-relativistic regime, one can compute correlation functions on the dual field theory side with the hope that they may help to constrain many of the correction terms. We will present a simple non-relativistic higher-spin/field theory duality proposal in Section \ref{sec:holo}.

%%%%%%%%%%%%%%%%%%%%%%%%%%%%%%%%%%%%%%%%%%%%%%%%%%%%%%%%%%%%%
\subsection{Some simple correlation functions}\label{sec:correlation-function}
Given that our bulk-to-boundary propagator has the same simple expression for all spins in the light-cone gauge, let us proceed and compute some simple correlation functions such as the 2-point and 3-point functions. Possible checkpoints for our computation will be those in \cite{Balasubramanian:2008dm,Volovich:2009yh,Fuertes:2009ex,Leigh:2009eb}. These work have previously considered field theories that admit gravitational duals on Schr\"{o}dinger background.

%%%%%%%%%%%%%%%%%%%%%%%%%%%%%%
\paragraph{2-point function:} The 2-point function can be obtained from the on-shell action of a free scalar field in the bulk. In particular, following the standard procedure in \cite{Balasubramanian:2008dm}, (see also Appendix \ref{app:2pt}), we obtain 
\begin{align}\label{eq:2-pt-function}
    \mathrm{(2pt)}\equiv \langle O(p_1)O(p_2)\rangle&=\delta^3(\vec{p}_1+\vec{p}_2)\lim_{\epsilon\rightarrow 0}\Bigg(\lim_{r\rightarrow \epsilon}\epsilon^{1-2\nu} \cK(p_1,r)\p_r \cK(p_2,r)\Bigg)\nn\\
    &=\delta^3(\vec{p}_1+\vec{p}_2)\frac{\Gamma(1-\nu)}{\Gamma(\nu)}|p_1|^{2\nu}\,.
\end{align}
One can check that upon Fourier transforming the above (two-point) functions, we can obtain a similar structure with the two-point functions 
in position space, cf.  \cite{Balasubramanian:2008dm,Fuertes:2009ex}. (See also Appendix \ref{app:2pt}.) 
%Note, however, that we need to do a Wick rotation $x^+\mapsto -i\tau$ to go to Euclidean space for the 2-pt function in position space to be well-defined. See Appendix \ref{app:2pt}.
Next, to extract Kubo-like quantities (see e.g. \cite{Nastase:2017cxp}), we define
\begin{align}
    \boldsymbol{\eta}[m]:=\frac{1}{\gamma}\mathrm{Im}\,(\mathrm{2pt})\,.
\end{align}
These quantities are only different by the mass $m_s$, which can be spin dependent. (We will elaborate this matter in the next section.) Now, in the static case,
\begin{align}\label{eq:kubo-response}
   \lim_{\gamma\rightarrow 0} \boldsymbol{\eta}[m]\sim \lim_{\gamma\rightarrow 0}\frac{1}{\gamma}(\mathrm{2pt})\Big|_{\rho=0}\gg 1\,,
\end{align}
which signals the existence of conserved higher-spin modes. As a result, it is tempting to say that the holographic dual of our non-relativistic massive chiral HSGRA may be a weakly coupled Laudau-Ginzburg theory, which describes a two-fluid system with a $\lambda$-point. This is in accordance with the strength of the 't Hooft coupling $\lambda\sim e^{\sigma}=e^{-\frac{1}{r}}\in (0,1)$, which indicates that our dual field theory is effectively free in the UV (where $r\rightarrow 0$), and becomes slightly more interacting in the IR (where $r\rightarrow \infty$).\footnote{Note that the radial coordinate $r$ may be interpreted as the RG scale in the boundary theory.}

%%%%%%%%%%%%%%%%%%%%%%%%%%%%%%
\paragraph{3-pt functions.} As alluded to above, we do not have complete information about the cubic vertices, cf. \eqref{eq:NR-cubic}. Nevertheless, we can still proceed with the part of the data available after compactification, which is $V_3[h_1,h_2,h_3]$. This portion of the three-point functions for spinning fields reads:\footnote{See also \cite{Henkel:1993sg} for previous work. Note that we drop the $\delta$ function on the support of boundary momentum conservation for simplicity.}
\begin{align}\label{eq:to-be-computed}
    (\mathrm{3\text{-}pt})'=\int dr dr_1dr_2dr_3\,\frac{\sqrt{r_1\,r_2\,r_3}}{\sqrt{|p_1||p_2||p_3|}}K_{\nu_1}(|p_1|r_1)K_{\nu_2}(|p_2|r_2)K_{\nu_3}(|p_3|r_3)V_3[h_1,h_2,h_3]\,.
\end{align}
Note that $\sqrt{\epsilon}K_{\nu}(|p|\epsilon)$ factors can be absorbed appropriately, as the integral over the radial coordinate $r$ is completely regular. 
In computing the above integral, we shall express two of the Bessel functions of the second kind $K_{\nu}(z)$ in terms of $J_{\nu}(z)$ via
\begin{align}
    K_{\nu}(z)=\frac{\pi}{2\sin(\pi\nu)}\big(I_{-\nu}(z)-I_{\nu}(z)\big)=\frac{\pi}{2\sin(\pi\nu)}\Big[i^{\nu}J_{-\nu}(iz)-i^{-\nu}J_{\nu}(iz)\Big]\,,
\end{align}
then subsequently apply the formula {\bf 6.578} in \cite{gradshteyn2014table}
\begin{align}\label{eq:rule1}
    &\int_0^{\infty}x^{\rho-1}J_{\lambda}(ax)J_{\mu}(bx)K_{\nu}(cx)dx\nn\\
    &=\frac{2^{\rho-2}a^{\lambda}b^{\mu}c^{-\rho-\lambda-\mu}}{\Gamma(\lambda+1)\Gamma(\mu+1)}\Gamma\Big(\frac{\rho+\lambda+\mu-\nu}{2}\Big)\Gamma\Big(\frac{\rho+\lambda+\mu+\nu}{2}\Big)\nonumber\\
    &\times F_4\Big(\frac{\rho+\lambda+\mu-\nu}{2},\frac{\rho+\lambda+\mu+\nu}{2};\lambda+1;\mu+1;-\frac{a^2}{c^2};-\frac{b^2}{c^2}\Big)\,,
\end{align}
\normalsize
where $F_4$ is an Appell function, to obtain parts of the 3-point functions.

Unfortunately, the final expressions do not admit a concise form. (We refer the reader to Appendix \ref{app:C} for a sketch of the computation.) This is expected since the underlying symmetry of the NR massive model is not as powerful as the one in massless chiral HSGRA. (See \cite{Skvortsov:2018uru} for 3-pt functions in exact $AdS_4$, which have very simple expressions.)

%%%%%%%%%%%%%%%%%%%%%%%%%%%%%%%%%%%%%%%%%%%%%%%%%%%%%%%%%%%%%
\subsection{Higher-spin interaction suppression}\label{sec:HS-breaking}
There are various reasons to believe that the interactions between massive higher-spin fields in the proposed theory should be suppressed. We will now show that if the masses are related to spins in an intricate way, higher-spin interactions will indeed be suppressed. Suppose momentum conservation in the $p^+$-direction holds even after compactification, i.e.
\begin{align}\label{eq:momentum-conservation}
    \sum_{a=1,2,3}\beta_a=0\mapsto \sum_{a=1,2,3}\sign(h_a)m_a(s_a)=0\,.
\end{align}
Then, since $m_a$ come from a relativistic system, we may assume the spin-mass relations, at $\tz=1$, follow the usual Regge trajectory. Namely,
\begin{align}\label{eq:mass-assumption}
    s=\alpha_0+\alpha'\,m^2\,,
\end{align}
where $\alpha_0$ is the intercept, and $\alpha'$ is the slope. (A somewhat similar idea that emphasizes on describing string theory as a broken phase of higher-spin gravity can be found in \cite{Metsaev:1999kb}.) 
We now want to propose a formula, which can capture the approximate change of the above mass-spin relation for $\tz\in[1,2]$, in the $c=\hbar=1$ unit. 

Let us make a rather strong assumption that massive spinning particles have some sort of bodies after doing the above null reduction to go to a non-relativistic regime. That is, they are composite and may be described by a spinning sphere or a spinning string of size $L$, aligning with the previous consideration in  \cite{takabayasi1979relativistic,takabayasi1974theory,kojima1979relativistic}. 
%This consideration is reasonable since the background vielbein \eqref{eq:vierbein} and the dilaton \eqref{eq:dilaton}, which couple to gauge fields, can easily spoil higher-spin symmetry as they ``\emph{slows}'' down relativistic higher-spin fields, effectively making them massive -- in the sense of moving slower than the speed of light. Thus, the resulting massive theory can describe a lower dimensional system via an act of compactification. 

Now, in the NR regime, we can imagine particles as some spinning sphere with fixed radius $L$ and the angular velocity $v$. Then, the total angular momentum (viewed as the ``classical spin'') reads very roughly as $J\sim m$. On the other hand, in the relativistic regime, we can view our field as an open string of size $L$ with some tension $T$. As the string is stressed, we can assume that the relativistic string carries an energy of $E\sim TL$. Then, the angular momentum scales as $J\sim p\frac{E}{T}\sim \frac{p^2}{T}$ where $E=\sqrt{m^2+p^2}$. 
This implies that in the relativistic regime, $J\sim m^{2}$. As a consequence, we can consider
\begin{align}
    s=\alpha_0+\alpha'm^{\frac{2}{\tz}}\,,
\end{align}
to be an approximate spin-mass relation that can interpolate between $\tz\in[1,2]$. (A more sophisticated way to obtain the above approximate formula can be found, e.g. in \cite{takabayasi1979relativistic,takabayasi1974theory,kojima1979relativistic}.) Observe that at $\tz=2$, and for sufficiently large $s$, 
\begin{align}
    m_s\,\propto \,s\,.
\end{align}
This implies that $\sum_ah_a=0$ for sufficiently large spins in $3d$ via \eqref{eq:momentum-conservation}. Thus, there will be no cubic interactions in this case, as the coupling constant $1/\Gamma(\sum_ah_a)$ will effectively trivialize cubic interactions in the large spin sector, leading to a suppression of higher-spin interactions. From the low-energy physics point of view, this is totally expected and acceptable.

%%%%%%%%%%%%%%%%%%%%%%%%%%%%%%%%%%%%%%%%%%%%%%%%%%%%%%%%%%%%%
\section{A holography conjecture}\label{sec:holo}

Given all the information of the bulk theory presented above, let us now give a discussion on what the dual theory on the boundary should be. 

As briefly discussed above, the ’t Hooft coupling $\lambda$ is related to the vev. of the dilaton via $\lambda\sim e^{-1/r}$. Thus, the boundary theory is expected to be asymptotically free at high energy, and weakly interacting in the IR.
%This suggests that $\lambda\in(0,1)$ and the boundary theory to start with should be a free or weakly coupled field theory before the Lifshitz dynamical exponential kicking in. 
These facts align with the duality between parity-violated higher-spin gravity and Chern-Simons matter theories proposed in \cite{Giombi:2011kc}. In the presence of torsion, it is expected that the theory on the resulting boundary theory should contain some massive gauge fields $A^{\mu}$ with certain null profile along $x^+$ to break Lorentz invariance. From the holography perspective, any change in these massive gauge fields will trigger the change of the bulk geometry.

Given these data points, we conjecture that there should be a duality between
\begin{itemize}
    \item[-] a NR massive chiral HSGRA in deformed $AdS_3$, and
    \item[-] a NR massive critical $2d$ chiral scalar theory obtained via compactification of the chiral sector of $3d$ Chern–Simons-matter theories, cf. \cite{Jain:2024bza,Metsaev:2018xip,Aharony:2024nqs, Chen:2025xlo}. This theory is anticipated to describe a two-fluid system (with a $\lambda$-point featuring a 2nd order phase transition) in a constrained space along the line of \cite{Son:2005rv,Geracie:2015drf}.
\end{itemize}
If correct, this conjecture should extend the statement of \cite{Klebanov:2002ja,Sezgin:2002rt} to the non-relativistic case. We shall study this conjecture in a companion work.

%%%%%%%%%%%%%%%%%%%%%%%%%%%%%%%%%%%%%%%%%%%%%%%%%%%%%%%%%%%%%
\section{Conclusion}\label{sec:diss}
In this work, we study the non-relativistic limit of massless chiral HSGRA in $AdS_4$ by introducing a Lifshitz dynamical exponential that drags the relativistic system to a lower dimensional NR chiral massive theory in a deformed $AdS_3$. We find that the resulting non-relativistic chiral massive theory is not topological and can have reasonable physics with higher-spin interactions suppressed in the large spin sector. We expect that the massive theory in this work can be used to describe a two-fluid system in one spatial dimension \cite{Son:2005rv,Geracie:2015drf}. This study will be done in a companion work.

Based on the fact that the Schr\"odinger algebra is a light-cone one, we have adapted our analysis along the settings of light-front formalism, cf. \cite{Metsaev:1999ui,Metsaev:2008fs,Metsaev:2018xip,Skvortsov:2018uru}. This turns out to be an advantage, as some observables can be computed quite easily. Nevertheless, we are not able to fix all the couplings, leaving the $3d$ chiral massive theory only partially determined. This stems from the fact that all Lorentz boosts generators either descend to Galilean ones or simply absent after Lifshitz scalings kick in and deform the geometry. 

If our conjectures regarding the duality between $(1+1)$-dimensional superfluid and $3d$ NR massive higher-spin gravity hold approximately, it may offer a possibility to study the superfluid-to-fluid transition from a gravitational perspective. Intriguingly, this aligns with the conjecture in the early 2000s by Klebanov and Polyakov that higher‑spin gravity might provide insight into classical critical phenomena of the $3d$ Wilson-Fisher $O(N)$ vector models \cite{Klebanov:2002ja}. Furthermore, as our bulk theory consists of complex massive higher-spin fields, it would be interesting to study the effects of temperature in chiral higher-spin gravity by considering blackbrane geometry (see e.g. \cite{Kovtun:2003wp,Bekaert:2011qd,Gutperle:2013oxa,Beccaria:2015iwa,Kolekar:2016pnr,Kolekar:2016yzg,Mukherjee:2017ynv} for previous work in this direction).

%%%%%%%%%%%%%%%%%%%%%%%%%%%%%%%%%%%%%%%%%%%%%%%%%%%%%%%%%%%%%
\section*{Acknowledgement}
We thank Dung Xuan Nguyen, Dam Thanh Son and Zhenya Skvortsov for various useful discussions. We also thank an anonymous PRD referee for many valuable
suggestions to improve our work. 
T.T thanks the ICISE for creating a fruitful environment for discussion during the Advanced Summer School in Quantum Field Theory and Quantum Gravity in Quy Nhon, Vietnam. D.M is partially supported by POSTECH Grant No. 4.0031175.01. D.M. also acknowledges support by the National Research Foundation of Korea (NRF) grant funded by the Korean government (MSIT) (No. 2022R1A2C1003182). T.T is supported by the National Research Foundation of Korea (NRF) grant funded by the Korean government (MSIT) (No. RS-2025-00564305)). A.M was supported by POSTECH BK21 Postdoctoral Fellowship during the initial part of the work. D.M and T.T were partially supported by the Young Scientist Training (YST) program at the Asia Pacific Center of Theoretical Physics (APCTP) through the Science and Technology Promotion Fund and Lottery Fund of the Korean Government, and also the Korean Local Governments
– Gyeongsangbuk-do Province and Pohang City.

%%%%%%%%%%%%%%%%%%%%%%%%%%%%%%%%%%%%%%%%

%%%%%%%%%%%%%%%%%%%%%%%%%%%%%%%%%%%%%%%%%%%%%%%%%%%%%%%%%%%%%

\appendix
%%%%%%%%%%%%%%%%%%%%%%%%%%%%%%%%%%%%%%%%%%%%%%%%%%%

%%%%%%%%%%%%%%%%%%%%%%%%%%%%%%%%%%%%%%%%%%%%%
\section{On the rescaled spinning fields}\label{app:B}
Let us elaborate how to reach the action \eqref{eq:Lagrangian-s} from covariant gauge. 
%Another effect that this rescaling brings into our construction is that all the mass parameters obtained via null-reduction are dimensionless.
Given that our framework involves complex scalar fields in the light-cone gauge, it may be unnecessary to employ Fronsdal field representations, see e.g. the reviews \cite{Didenko:2014dwa,Ponomarev:2022vjb}, which are doubly-traceless and primarily suited for maximally symmetric background. Rather, we can consider chiral field representations, cf. \cite{Krasnov:2021nsq}, which have a close connection to the light-front approach.

There exists a well-known action that remains valid for any self-dual background -- a broader class of backgrounds whose the anti-self-dual component of the Weyl tensor vanishes -- which has the following simple form:
\begin{align}\label{eq:action-SD}
    S_2[s]=\int d^4x\sqrt{-g}B^{\alpha(2s)}\nabla_{\alpha}{}^{\dot\alpha}A_{\alpha(2s-1)\,\dot\alpha}\,,\qquad \alpha=1,2\,,\quad \dot\alpha=\dot 1,\dot 2\,.
\end{align}
Here, $A_{\alpha(2s-1)\,\dot\alpha}$ denotes the positive-helicity spin-$s$ field, and $B^{\alpha(2s)}$ denotes the negative-helicity spin-$s$ field. The action is invariant under $\delta A_{\alpha(2s-1)\,\dot\alpha}=\nabla_{\alpha\dot\alpha}\xi_{\alpha(2s-2)}$.

To project \eqref{eq:action-SD} to the light-cone gauge, we shall focus on the equations of motion of the positive-helicity fields. In exact $AdS_4$, $\nabla$ acts on $A^{\alpha(2s-1)\,\dot\alpha}$ as
\begin{align}
    \nabla^{\alpha}{}_{\dot\alpha}A^{\alpha(2s-1)\,\dot\alpha}=\p^{\alpha}{}_{\dot\alpha}A^{\alpha(2s-1)\,\dot\alpha}+(2s-1)\Pi^{\alpha\ \ \alpha}_{\ \ \gamma\ \ \dot\alpha}A^{\alpha(2s-2)\gamma\,\dot\alpha}+\Pi^{\dot\alpha \ \ \alpha}_{\ \ \dot\gamma \ \ \dot\alpha}A^{\alpha(2s-1)\,\dot\gamma}\,,
\end{align}
where we have used the following notations: 
\begin{align}
      \Pi^{\alpha\ \ \alpha}_{\ \ \gamma\ \ \dot\alpha}=(\varpi_{\mu})^{\alpha}{}_{\gamma}(\tE^{\mu})^{\alpha}{}_{\dot\alpha}\,,\qquad 
     \Pi^{\dot\alpha \ \ \alpha}_{\ \ \dot\gamma \ \ \dot\alpha}=(\varpi_{\mu})^{\dot\alpha}{}_{\dot\gamma}(\tE^{\mu})^{\alpha}{}_{\dot\alpha}\,,
\end{align}
and the convention for symmetrization as $v^{\alpha}u^{\alpha}=\frac{1}{2}(v^{\alpha_1}u^{\alpha_2}+v^{\alpha_2}u^{\alpha_1})$, etc. Note that $\tE^{\alpha\dot\alpha}$ are the inverse vielbein of exact $AdS_4$, whose line element is \eqref{eq:exact-AdS}. Moreover, $\varpi_{\alpha\beta}$ and $\varpi_{\dot\alpha\dot\beta}$ are the anti-self-dual (ASD) and self-dual (SD) components of the spin-connection $\varpi^{ab}=-\varpi^{ba}$ in vectorial language. Their explicit forms can be computed via the usual Cartan structures equations, see for instance \cite{Bolotin:1999fa}, (we set $L=1$ for simplicity here)
\begin{align}\label{eq:spin-connection-AdS}
    \varpi_{\alpha\beta}=+\frac{i}{\sqrt{2}}\frac{\sigma_{\alpha\beta}}{r}\,,\qquad \varpi_{\dot\alpha\dot\beta}=-\frac{i}{\sqrt{2}}\frac{\sigma_{\dot\alpha\dot\beta}}{r}\,.
\end{align}
A short computation shows that
\begin{align}
    \nabla^{\alpha}{}_{\dot\alpha}A^{\alpha(2s-1)\,\dot\alpha}=\p^{\alpha}{}_{\dot\alpha}A^{\alpha(2s-1)\,\dot\alpha}-i\frac{(s+1)}{\sqrt{2}}A^{\alpha(2s)}\,.
\end{align}
At this stage, we can impose the light-cone gauge where $A^{\alpha(2s-2)0\,\dot 0}=0$. Then, the above equation results in a set of two equations\footnote{
In the spinorial notation, a point $x^{\alpha\dot\alpha}$ on the tangent space can be expressed in light-cone coordinate as 
\begin{align}
    x^{\alpha\dot\alpha}=\frac{1}{\sqrt{2}}\begin{pmatrix}
    x^2+x^0 & x^1-ir\\
    x^1+ir & x^2-x^0
    \end{pmatrix}=\begin{pmatrix}
    x^+ & x^R\\
    x^L & -x^-
    \end{pmatrix}\quad \leftrightarrow\quad \p^{\alpha\dot\alpha}=\begin{pmatrix}
        \p^+ & \p^R\\
        \p^L & -\p^-
    \end{pmatrix}\,,
\end{align}
where the basis for Pauli matrices associated with a spacetime point $x^{\mu}=(x^0,x^1,x^2,r)$ is 
\begin{align}
    \sigma_{\mu}^{\alpha\dot\alpha}=\frac{1}{\sqrt{2}}(\onebb,\sigma_1,\sigma_3,\sigma_2)\,.
\end{align}
Our normalization is such that $\sigma^{\alpha\dot\alpha}\sigma^{\beta\dot\beta}=\epsilon^{\alpha\beta}\epsilon^{\dot\alpha\dot\beta}$, and $\p^{\alpha\dot\alpha}x^{\beta\dot\beta}=\epsilon^{\alpha\beta}\epsilon^{\dot\alpha\dot\beta}$. The derives are given by
\begin{align}
    \p^L=\frac{\p_1+i\p_r}{\sqrt{2}}\,,\qquad \p^R=\frac{\p_1-i\p_r}{\sqrt{2}}\,,\qquad \p^Lx^R=1\,,\quad \p^Rx^L=1\,.
\end{align}
Denoting  $\epsilon^{\alpha\beta},\epsilon^{\dot\alpha\dot\beta}$ as $SL(2,\C)$ invariance tensors, we raise and lower the spinorial indices as
\begin{align}
    X^\alpha\epsilon_{\alpha\beta}=X_{\beta}\,,\quad \epsilon^{\alpha\beta}X_\beta=X^\alpha\,,%\qquad X^1\epsilon_{10}=-X_0\,,\quad X^0\epsilon_{01}=X_1\quad \Rightarrow 
    \quad X_0=-X^1\,,\ X_1=X^0\,.
\end{align}} 
\begin{subequations}\label{eq:LC-solve-for}
    \begin{align}
        F^{1(2s-1)0}&=-\p^L\varphi_{+s}+\p^+\Xi_{+s}-i\frac{(s+1)}{\sqrt{2}}\varphi_{+s}=0\,,\\
        F^{1(2s)}&=+\p^-\varphi_{+s}+\p^R\Xi_{+s}-i\frac{(s+1)}{\sqrt{2}}\Xi_{+s}=0\,,
    \end{align}
\end{subequations}
where $\Xi_{+s}$ is an auxiliary field that needs to be solved for. Observe that upon a simple rescaling\footnote{Note that $s+1$ happens to be the conformal dimensions on the unitary bound for all massless higher-spin fields in $AdS_4$, see e.g. \cite{Metsaev:1997nj}.}
\begin{align}
    \varphi_{+s}\mapsto r^{s+1}\varphi_{+s}\,,\qquad \Xi_{+s}\mapsto r^{s+1}\Xi_{+s}\,,
\end{align}
we can reduce \eqref{eq:LC-solve-for} to:
\begin{align}
    \p^-\varphi_{+s}+\frac{\p^R\p^L}{\p^+}\varphi_{+s}=0\,,
\end{align}
after solving for $\Xi_{+s}$ in terms of $\varphi_{+s}$. Treating $B_{1(2s)}=\varphi_{-s}$, and making a suitable rescaling, we obtain
\begin{align}
    S_2[s]=\int d^4x\,\varphi_{-s}\,\Box\,\varphi_{+s}\,,
\end{align}
in what we will call Metsaev frame \cite{Metsaev:1999ui} -- a choice of gauge up to field redefinition that makes the kinetic terms for massless spinning fields in $AdS_4$ look similar to the ones in flat space.

%Let us elaborate on the rescaling of fields in \cite{Metsaev:1999ui}, which effectively make all spinning fields have the same conformal dimensions. 
%As already mentioned before, the central motivation for this change is to ensure a part of the kinetic operator mimics the Laplacian of flat spacetime, leading to simpler final expressions of observables. 
%%%%%%%%%%%%%%%%%%%%%%%%%%%%%%%%%%%%%%%%%%%%%%%%
To this end, we can add the term $\upsilon^2\int \varphi_{-s}r^{2-2\tz}(\partial_-)^2\varphi_{+s}$ into the above action, to obtain the correct Hamiltonian $P^-$, cf. \eqref{eq:Hamiltonian}, for all free higher-spin fields. This is acceptable due to the fact that we are not obligated to fulfill gauge invariance as the Lifshitz deformation \eqref{eq:M-scale} drags our $4d$ massless theory to a $3d$  massive one.

%%%%%%%%%%%%%%%%%%%%%%%%%%%%%%%%%%%%%%%%%%%%%%%%%%%%%%%%%%%

%%%%%%%%%%%%%%%%%%%%%%%%%%%%%%%%%%%%%%%%%%%%%%%%%%%%%%%%%%%%%%%%%%%%
%%%%%%%%%%%%%%%%%%%%%%%%%%%%%%%%%%%%%%%%%%%%%%%%%%%%%%%%%%%%%%%%%%%%%%%%%%%%%%%%%%%%%%%%
\section{Computations of correlation functions}\label{app:C}

This appendix contains some information about the computations for the 2-point and 3-point functions.

%%%%%%%%%%%%%%%%%%%%%%%%%%%%%%%%%%%%%%%%%%%%
\subsection{Calculation of 2-point function}\label{app:2pt}

%From our previously derived expression in \eqref{eq:bulk-to-bdy-prop}, the scalar field in the bulk can be expressed in terms of its boundary value $\varphi_{\rm bdy}(p,\epsilon)$ (at some cut-off UV surface $r=\epsilon$) as
%\begin{equation}
%    \varphi(x^+, x^{-}, x^1, r)= \frac{1}{(2\pi)^{3/2}}\int d^3p~ e^{i(\gamma x^+ + \beta x^{-}+\rho x^1)}\cK(p,r)\varphi_{\rm bdy}(p,\epsilon)\ . 
%\end{equation}
The two-point function of the operator, dual of the \emph{rescaled} scalar field follows from the on-shell action that takes the form
%\begin{equation}
%    S_{\rm bdy}=\int_{\partial \mathcal{M}}d^3x\ \sqrt{\gamma}g^{r\mu}\varphi \partial_{\mu}\varphi = \frac{1}{(2\pi)^3L^2}\int_{\partial \mathcal{M}} d^3p\ \varphi_{\rm bdy}(p,\epsilon)\mathcal{F}(p)\varphi_{\rm bdy}(-p,\epsilon)\ ,
%\end{equation}
%\tnote{I think it should be}
\begin{equation}
    S_{\rm bdy}=\frac{1}{L}\int_{\partial \mathcal{M}}d^3x\,\varphi \partial_{r}\varphi = \int d^3p\ \varphi_{\rm bdy}(p,\epsilon)\mathcal{F}(p)\varphi_{\rm bdy}(-p,\epsilon)\ ,
\end{equation}
%\tnote{as we are working with the action \eqref{eq:free-scalar-NR}. Note that we have rescaled $L$ away.}
where
%\begin{equation}
%    \mathcal{F}(p)=\lim_{r \to \epsilon}\frac{1}{r}\cK(p,r)\partial_r \cK(p,r)
%\end{equation}
%\tnote{I think it should be}
\begin{equation}\label{eq:Flux}
    \mathcal{F}(p)=\lim_{r \to \epsilon}\frac{1}{L}\cK(p,r)\partial_r \cK(p,r)
\end{equation}
$\cK(p,r)$ was introduced earlier in \eqref{eq:phiansatz}. Finally, the 2-point correlator of the dual operator in momentum space can be written as
\begin{equation}
    \langle O(p_1)O(p_2)\rangle = %\frac{1}{L^2}
    \delta^3(p_1+p_2)\left(\mathcal{F}(p_1)+\mathcal{F}(-p_1) \right)
\end{equation}
%\tnote{We are working with the rescaled field $\varphi\rightarrow r/L \varphi$, I think there is no $L$, see the dimension of variables in (4.3). $\nu$ is dimensionless, so it is a number. Everything else are functions of momentum. There should be no other dimensionful parameters after redefining fields.}
Now, to evaluate the above near the boundary, we require the appropriate power series expansion for the modified Bessel function
\begin{equation}\label{eq:K-Bessels-sub}
    K_{\nu}(x)= x^{-\nu}(a_0+a_1x^2+a_2x^4+ \cdots)+x^{\nu}(b_0+b_1x^2+b_2x^4+\ldots)\,,
\end{equation}
where
\begin{align}
    a_0=2^{\nu-1}\Gamma(\nu)\,,\qquad b_0=-2^{-\nu-1}\Gamma(1-\nu)/\nu\,.
\end{align}
Note that all other higher coefficients will not play important roles here. Substituting \eqref{eq:K-Bessels-sub} to \eqref{eq:Flux} and taking the limit $r\rightarrow \epsilon$ carefully, we obtain \eqref{eq:2-pt-function}.

\paragraph{Position space representation.} Following e.g. Appendix B in \cite{Fuertes:2009ex}, let us rewrite the 2-pt function \eqref{eq:2-pt-function} in position space by considering 
\begin{align}
    G(x^+,x^1,x^-)=\int \frac{d\beta}{2\pi}e^{i\beta x^-}\int\frac{d\gamma d\rho}{(2\pi)^2}|p|^{2\nu}e^{i(\gamma x^++\rho x^1)}\,,\qquad \nu>0\,,
\end{align}
where $p^2=2\beta\gamma+\rho^2$. Taking $\beta\neq 0$ where $\beta\in \mathbb{R}$ ($\beta$ becomes the physical non-relativistic mass after dimensional compactification along the $x^-$ direction, which can be $\pm m$ for $m>0$), the above can be reduced to
\begin{align}
    G(x^+,x^1,x^-)&=\int \frac{d\beta}{2\pi}e^{i\beta x^-}\frac{(2\beta)^{\nu}}{\pi}\int_0^{\infty}d\rho \cos(\rho x^1)\int \frac{d\gamma}{2\pi}e^{i\gamma x^+}\Big(\gamma+\frac{\rho^2}{2\beta}\Big)^{\nu}\,.
\end{align}
First, we focus on the $\gamma$ integral. Following \cite{Fuertes:2009ex}, we first perform a shift $\gamma \to \gamma - \frac{\rho^2}{2\beta}$ in the integration variable which results in  
\begin{equation}
    G(x^{+},x^1,x^{-})=\int \frac{d\beta}{2\pi}e^{i\beta x^-}\frac{(2\beta)^{\nu}}{\pi}\int_0^{\infty}d\rho \cos(\rho x^1)e^{-\frac{i\rho^2x^{+}}{2\beta}}\int_{-\infty}^{+\infty} \frac{d\gamma}{2\pi}e^{i\gamma x^+}\gamma^{\nu}\ .
\end{equation}
It is then convenient to choose the branch cut along the positive imaginary axis connecting the branch points at $\gamma=0$ and $\gamma=i\infty$ 
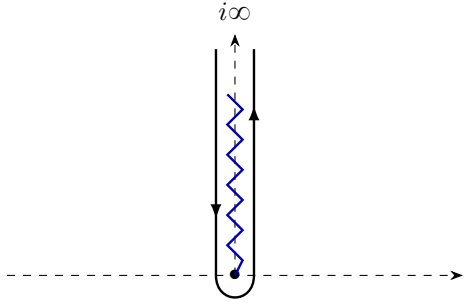
\begin{center}
    \begin{tikzpicture}[font=\sffamily\small]
        %\draw[thin,dotted] (-6,0) grid (6,6);
        \draw[->,-Stealth,dashed] (-3,0) to (3,0);
        \draw[->,-Stealth,dashed] (0,0) to (0,3.2);
        %%%%%%%%%%%
        \node at (0,0) (bp) {$\bullet$};
        \node at (0,3.5) (x) {$i\infty$}; 
        %%%%%%%%%%%%%%
        \draw[line width=0.9pt,middlearrow={latex}] (-0.25,3) to (-0.25,0.0);
        \draw[line width=0.9pt,out=-90,in=-90,looseness=2] (-0.25,0) to (0.25,0);
        \draw[line width=0.9pt,middlearrow={latex}] (0.25,0) to (0.25,3);
        \draw[line width=0.9pt,color=blue!70!black!100,looseness=1.1] (0,0) -- (0.1,0.2) -- (-0.1,0.4) -- (0.1,0.6) -- (-0.1,0.8) -- (0.1,1) -- (-0.1,1.2) -- (0.1,1.4) -- (-0.1,1.6) -- (0.1,1.8) -- (-0.1,2) -- (0.1,2.2) -- (-0.1,2.4);
        %%%%%%%%%%%%%%%%%%%%%%%%%%%%%%%%%%%%%%%%%%
    \end{tikzpicture}
\end{center}
Upon parametrizing $\gamma$ as $\gamma=\pm \epsilon + i\chi$ along the contours running counterclockwise near the branch cut, and using the Cauchy's residue theorem, we obtain
\begin{equation}
\label{eq:gamma-integral}
   \int_{-\infty}^{\infty} \frac{d\gamma}{2\pi}e^{i\gamma x^{+}}\gamma^{\nu} = i^{\nu+1}(e^{2\pi i \nu}-1) \int_0^{\infty}d\chi\ \chi^{\nu}e^{-\chi x^{+}} = i^{\nu}e^{i\pi \nu}\frac{(x^{+})^{-\nu-1}}{\Gamma(-\nu)}\theta(x^{+})\ .
\end{equation}
At this stage let us replace $x^+=t$, $x^1=x$ and $x^-=\xi$ to ease the notations. Then, our 2-pt function reads
\begin{align}
    G(t,x,\xi)&=\theta(t)\int \frac{d\beta}{2\pi}e^{i\beta \xi}\frac{(2\beta)^{\nu}}{\pi}\int_0^{\infty}d\rho \cos(\rho x^1)e^{-i\frac{\rho^2 t}{2\beta}}\times i^{\nu}e^{i\pi \nu}\frac{1}{\Gamma(-\nu)}\frac{1}{t^{\nu+1}}\,.
\end{align}
Here, the factor $\theta(t)$ in front encapsulates the fact that this is the \emph{retarded} propagator.
Our next task is to perform the $\rho$-integral, which results in 
%which will lead to the position space representation of the correlator in the effective 2-dimensional $x^{-}$-compactified Schr\"{o}dinger field theory. So,
\begin{equation}
\begin{aligned}
    G(t,x,\xi)&=\theta(t)\int \frac{d\beta}{2\pi}e^{i\beta \xi}\frac{(-2i\beta)^{\nu}}{\pi \Gamma(-\nu)}\frac{1}{t^{\nu+1}}\int_0^{\infty}d\rho\ \cos(\rho x^1)e^{-\frac{i\rho^2t}{2\beta}}\\
    &=\theta(t)\int \frac{d\beta}{2\pi}e^{i\beta \xi}\frac{(-2i\beta)^{\nu}}{\sqrt{2\pi}\, \Gamma(-\nu)}\frac{\sqrt{\beta}}{t^{\nu+\frac{3}{2}}}e^{-i\frac{\pi}{4}}e^{\frac{ix^2\beta}{2t}}\ .
\end{aligned}
\end{equation}
Note that the above integral over $\rho$ is convergent provided Im$\left(\frac{t}{\beta}\right)<0$. 
%\textcolor{red}{This is indeed natural and consistent with \cite{Son:2008ye} where the conjugate momentum in the compactified theory should be identified with the non-relativistic mass i.e. $\frac{\partial}{\partial x^{-}}\sim \beta \sim im$. }
%\textcolor{red}{We differ by a sign as compared to \cite{Son:2008ye}, due to our definition of light-cone coordinate, $x^{-}$ which has an overall negative sign as compared to the one in \cite{Son:2008ye}.}

The two-point function of the effective $2d$ non-relativistic theory is then simply proportional to the integrand of the $\beta$-integral after stripping off the measure $\frac{e^{i\beta \xi}}{2\pi}$ and identifying $\beta=im$. The reason is that from a $2d$ perspective $\xi$ is an unphysical scaleless dimension. As such, we do \emph{not} need to integrate over $\beta$ if we want to obtain the expression of the 2-pt function in $2d$. Furthermore, as noted in \cite{Fuertes:2009ex}, the extraction of $2$ dimensional correlation functions in position space starting from a $3d$ expression in momentum space only makes sense in Euclidean signature. This suggests us to rotate $\beta=im$ and $\xi=-i\xi$, leading to the expected result 
\begin{equation}
\label{eq:rho-integral}
    G_{2d}(t,x) = \theta(t)\frac{2^{\nu-\frac{1}{2}}m^{\nu +\frac{1}{2}}}{\sqrt{\pi}\Gamma(-\nu)}\times\frac{1}{t^{\nu+3/2}}e^{-\frac{mx^2}{2t}}\,.
\end{equation}
The complete two-point function in position space can be obtained by multiplying the above with the $\Gamma(1-\nu)/\Gamma(-\nu)$ factor. We obtain
\begin{align}\label{eq:3d-result}
    (2pt)_{NR}=-2^{\nu}\theta(t)\frac{1}{(2\pi)^{1/2}}\frac{\nu}{\Gamma(\nu)}\frac{m^{\nu+\frac{1}{2}}}{t^{\nu+\frac{3}{2}}}\exp\Big(-\frac{m x^2}{2t}\Big)\,,
\end{align}
which has similar structure with the 2-pt function in \cite{Balasubramanian:2008dm,Fuertes:2009ex,Volovich:2009yh,Leigh:2009eb}. Note that the above makes sense only if $m>0$, which is natural from the non-relativistic view point.

%\tnote{Had we worked in Lorentzian time, the above would have oscillating behavior. The case where $\beta<0$ require $\tau>0$ and can be done similarly, we leave it as an exercise.}

Let us finish this appendix by giving the exact form of the position space correlator in the dual 3-dimensional theory by doing the $\beta$-integral
\begin{equation}
 \begin{aligned}
    G(t,x,\xi)&=\theta(t)\frac{(-1)^{\nu}2^{{\nu}-1/2}i^{\nu-1/2}}{\pi^{1/2}\Gamma(-\nu)}\frac{1}{t^{\nu+\frac{3}{2}}}\int_{-\infty}^{+\infty}\frac{d\beta}{2\pi}\ e^{i\beta \left( \xi+\frac{x^2}{2t}\right)}\beta^{\nu+\frac{1}{2}}\\
&=\theta(t)\theta\Big(\xi+\frac{x^2}{2t}\Big)\frac{(-1)^{\nu}2^{{\nu}-1/2}i^{\nu-1/2}}{\pi^{1/2}\Gamma(-\nu)}\frac{1}{t^{\nu+\frac{3}{2}}}\times \frac{i^{\nu+\frac{1}{2}}e^{i\pi\left(\nu+\frac{1}{2}\right)}}{\Gamma\left(-\nu -\frac{1}{2} \right)}\frac{(2t)^{\nu+\frac{3}{2}}}{(2t\xi+x^2)^{\nu+\frac{3}{2}}}\\
&=\theta(t)\theta\big(2t\xi+x^2\big)\frac{(-1)^{\nu+\frac{1}{2}}2^{2\nu+1}}{\sqrt{\pi}\Gamma(-\nu)\Gamma\left(-\nu-\frac{1}{2}\right)}\frac{1}{(2t\xi+x^2)^{\nu+\frac{3}{2}}}\ ,
 \end{aligned}   
\end{equation}
where we have used \eqref{eq:gamma-integral} to reach the second line. Note that $(2t\xi+x^2)=(-x_0^2+x_2^2+x_1^2)$ is precisely the distance between two points in $3d$ when we write light-cone coordinates in Cartesian ones, cf. \eqref{eq:exact-AdS}. Notice that the theta functions appearing in the above expression which automatically encodes the retarded nature of the propagator. Furthermore, as $t=x^+$ and $\xi=x^-$ are related by an overall rotation, one expects the final result to be symmetric in $x^{+}$ and $x^{-}$, which we happily see in the above expression. To this end, we write the three-dimensional 2-point function in position space as
\begin{align}
    (2pt)_{3d}=-(-1)^{\nu+\frac{1}{2}}2^{2\nu+1}\theta(t)\theta(2t\xi+x^2)\frac{\nu }{\Gamma(\nu)\Gamma\big(-\nu-\frac{1}{2}\big)}\frac{1}{(2t\xi+x^2)^{\nu+\frac{3}{2}}}\,.
\end{align}
Notice that the above 2-pt function is real as expected in non-relativistic physics.

%%%%%%%%%%%%%%%%%%%%%%%%%%%%%%%%%%%%%%%%%%%%%%%%%%%%%%%%%%%%%%%%%%%%%%%%%%%%%%%%%%%%%%%%
\subsection{Calculation of 3-point functions}
This appendix computes some simple 3-point functions from the portion of the 3-point vertices that we know from \eqref{eq:to-be-computed}. Recall that the propagator is
\begin{align}
    \cK(p,r)=\frac{1}{|p|}\frac{\sqrt{r}K_{\nu}(|p|r)}{\sqrt{\epsilon}K_{\nu}(|p|\epsilon)}\,,\quad \nu=\frac{1}{2}\sqrt{1+4\upsilon^2m^2}\,.
\end{align}
Expressing 
\begin{align}
    K_{\nu}(|p|r)=\frac{\pi}{2\sin(\pi\nu)}\big(I_{-\nu}(|p|r)-I_{\nu}(|p|r)\big)=\frac{\pi}{2\sin(\pi\nu)}\Big[i^{\nu}J_{-\nu}(i|p|r)-i^{-\nu}J_{\nu}(i|p|r)\Big]\,,
\end{align}
and absorbing $\sqrt{\epsilon}K_{\nu}(|p|\epsilon)$ factors in the definition of the portion of the 3-point function that we are computing, we can rewrite \eqref{eq:to-be-computed} as
\small
\begin{align}\label{eq:to-be-computed-1}
    \mathrm{3pt}'&=\int dr dr_1dr_2dr_3\,\frac{\sqrt{r_1r_2r_3}}{\sqrt{|p_1||p_2||p_3|}}K_{\nu_1}(|p_1|r_1)K_{\nu_2}(|p_2|r_2)K_{\nu_3}(|p_3|r_3)V_3[h_1,h_2,h_3]\nn\\
    &=\frac{1}{\sqrt{|p_1||p_2||p_3|}}\int dr dr_1dr_2dr_3\frac{\sqrt{r_1r_2r_3}i^{-\nu_1-\nu_2}}{4\sin(\pi\nu_1)\sin(\pi\nu_2)}\times\nn\\
    &\qquad \times\Big(i^{2\nu_1}J_{-\nu_1}(ip_1r_1)-J_{\nu_1}(ip_1r_1)\Big)\Big(i^{2\nu_2}J_{-\nu_2}(ip_2r_2)-J_{\nu_2}(ip_2r_2)\Big)K_{\nu_3}(p_3r_3)V_3[h_1,h_2,h_3]
\end{align}
\normalsize
Substituing \eqref{eq:V_3} and doing integrations by part with the note that
\begin{subequations}\label{eq:non-linear-der}
    \begin{align}
    \p_rJ_{\nu}(i|p|\,r)&=-\frac{i |p|}{2}\Big(J_{\nu+1}(i|p|\,r)-J_{\nu-1}(i|p|\,r)\Big)\,,\\
    \p_rK_{\nu}(|p|\,r)&=-\frac{|p|}{2}\Big(K_{\nu+1}(|p|\,r)+K_{\nu-1}(|p|\,r)\Big)\,,
\end{align}
\end{subequations}
we will get some integrals of the form
\small
\begin{align}\label{eq:replacement-rule}
    &\int_0^{\infty}r^{\rho-1}J_{\lambda}(i|p_1|r)J_{\mu}(i|p_2|r)K_{\nu}(|p_3|r)dr\nn\\
    &=\frac{2^{\rho-2}(i|p_1|)^{\lambda}(i|p_2|)^{\mu}(|p_3|)^{-\rho-\lambda-\mu}}{\Gamma(\lambda+1)\Gamma(\mu+1)}\Gamma\Big(\frac{\rho+\lambda+\mu-\nu}{2}\Big)\Gamma\Big(\frac{\rho+\lambda+\mu+\nu}{2}\Big)\nonumber\\
    &\times F_4\Big(\frac{\rho+\lambda+\mu-\nu}{2},\frac{\rho+\lambda+\mu+\nu}{2};\lambda+1;\mu+1;\frac{p_1^2}{p_3^2};\frac{p_2^2}{p_3^2}\Big)\,.
\end{align}
\normalsize
Notice that each time $\p_r$ acts on $J_{\nu}(i|p|r)$ or $K_{\nu}(|p|r)$, it will generate an extra term. Thus, as the number of the derivatives $\PP^L$ in the vertices go higher, the results will be more complicated but nevertheless controllable. The 3-pt functions with the portion that we know read
\begin{align}
    3pt'&=\frac{1}{\sqrt{|p_1||p_2||p_3|}}\int dr dr_1dr_2dr_3\frac{\sqrt{r_1r_2r_3}i^{-\nu_1-\nu_2}}{4\sin(\pi\nu_1)\sin(\pi\nu_2)}\nn\\
    &\qquad \times\Big(i^{2\nu_1}J_{-\nu_1}(i|p_1|r_1)-J_{\nu_1}(i|p_1|r_1)\Big)\Big(i^{2\nu_2}J_{-\nu_2}(i|p_2|r_2)-J_{\nu_2}(i|p_2|r_2)\Big)K_{\nu_3}(|p_3|r_3)\nn\\
    &\qquad \times\sum_{n=0}^{\tH-1}\frac{(r\PP^L)^{\tH-n}}{\,m_1m_2m_3}\cQ^n\prod_{a=1}^3\delta(r-r_a)\,,
\end{align}
where $\cQ$ is $r$ independent, and
\begin{align}
    \PP^L=\frac{1}{3\sqrt{2}}\sum_{a=1}^3\check{m}_a(\rho_a+\p_{r_a})\,.
\end{align}
Now, we can integrate by part, using \eqref{eq:non-linear-der} to generate the final form of the integral before substituing $r_{1,2,3}\mapsto r$ by integration over $\delta$-functions $\delta(r-r_a)$. After the dust settled, one can subsequently use \eqref{eq:replacement-rule} to get to the final expression. We refrain ourselves from exhibiting the final result since even with the total helicity $\tH=1$, we get 44 integrals of the form \eqref{eq:replacement-rule}. Since there are in general no constraints between $\nu_{1,2,3}$, the final result will be complicated.
%%%%%%%%%%%%%%%%%%%%%%%%%%%%%%%%%%%%%%%

%\newpage
\bibliography{Ref_NRDuality}
\bibliographystyle{JHEP}

\end{document}